\documentstyle{article}[12pt]

\def\dd{\mbox{d}}

\def\O{\Omega}
\def\o{\omega}

\def\a{\alpha}
\def\b{\beta}
\def\d{\delta}

\def\g{\gamma}
\def\G{\Gamma}
\def\e{\epsilon}

\def\f{\phi}
\def\F{\Phi}

\def\k{\kappa}

\def\m{\mu}
\def\n{\nu}

\def\o{\omega}
\def\p{\pi}
\def\r{\rho}
\def\t{\tau}

\def\pa{\partial}

\def\ta{\tilde{a}}

\def\ch{{\cal H}}

\def\cl{{\cal L}}

\def\cw{{\cal W}}

\renewcommand{\@}[1]{\sqrt{#1}}

\renewcommand{\le}[1]{\label{#1}\end{eqnarray}}
\newcommand{\be}{\begin{equation}}
\newcommand{\ee}{\end{equation}}
\newcommand{\bea}{\begin{eqnarray}}
\newcommand{\eea}{\end{eqnarray}}
\newcommand{\nn}{\nonumber}

\newcommand{\eq}[1]{(\ref{#1})}
\def\nn{\nonumber\\}

\def\ffract#1#2{\raise .35 em\hbox{$\scriptstyle#1$}\kern-.25em/
\kern-.2em\lower .22 em \hbox{$\scriptstyle#2$}}

\def\na{\nabla}
\def\half{{1\over2}\,}
\def\nonu{\nonumber \\{}}

\def\nonu{\nonumber \\{}}
\def\a{\alpha}
\def\b{\beta}
\def\c{\gamma}
\def\d{\delta}
\def\e{\epsilon}           
\def\f{\phi}               
\def\g{\gamma}

\def\k{\kappa}                    

\def\m{\mu}
\def\n{\nu}
\def\o{\omega}  
\def\p{\pi}                
\def\r{\rho}                                     
\def\t{\tau}

\def\x{\xi}

\def\F{\Phi}
\def\G{\Gamma}

\def\O{\Omega}

\def\S{\Sigma}

\begin{document}

\large
\null\vskip-24pt
\begin{flushright}
SWAT/310\\
UCLA/03/TEP/3\\
PUPT-2074 \\
ITFA-2003-06 \\
{\tt hep-th/0302136}
\end{flushright}
\begin{center}
\vskip .9truecm
{\Large\bf
On $\a'$-corrections to D-brane solutions}
\vskip 1truecm
{\large\bf Sebastian de Haro\footnote{
e-mail: {\tt sebas@physics.ucla.edu}},
{\large\bf Annamaria Sinkovics\footnote{e-mail: {\tt
sinkovic@science.uva.nl}}}
and Kostas Skenderis\footnote{
e-mail: {\tt skenderi@science.uva.nl}}}
\\
\vskip 1truecm
${}^{1}$ {\it University of California, Los Angeles\\
Los Angeles, CA 90095,USA}
\vskip 1truemm
${}^{2}$ {\it Department of Physics,
University of Wales Swansea\\
Singleton Park, Swansea, SA2 8PP, UK} \\
\vskip 1truemm
${}^{2}$ \  ${}^{3}${\it Institute for Theoretical Physics,
University of Amsterdam\\
Valckenierstraat 65, 1018XE Amsterdam, The Netherlands} \\
\vskip 1truemm
${}^{3}$ {\it Physics Department,
Princeton University \\
Princeton, NJ 08544, USA}
\end{center}
\vskip .5truecm
\begin{center}
{\bf  Abstract}
\end{center}
We discuss the computation of the leading corrections to D-brane solutions
due to higher derivative terms in the corresponding
low energy effective action.
We develop several alternative methods for analyzing the problem.
In particular, we derive an effective one-dimensional action
from which the field equations for spherically symmetric
two-block brane solutions can be derived, show how to obtain
first order equations, and  discuss a few other approaches.
We integrate the equations for extremal branes and obtain the
corrections in terms of integrals of the evaluation of the
higher derivative terms on the lowest order solution.
To obtain completely explicit results one would need to
know all leading higher derivative corrections which at present
are not available. One of the known higher derivative terms is
the $R^4$ term, and we obtain the corrections
to the D3 brane solution  due to this term alone.
We note, however, that (unknown at present) higher
terms depending on $F_5$ are expected to modify our result.
We analyze the thermodynamics
of brane solutions when such quantum corrections are present.
We find that the $R^4$ term induces a correction to the
tension and the electric potential of the D3 brane but not to its charge,
and the tension is still proportional to the
electric potential times  the charge. In the near-horizon
limit the corrected solution becomes
$AdS_5 \times S^5$ with the same cosmological
constant as the lowest order solution
but a different value of the (constant) dilaton.


\newpage

\section{Introduction} \label{intro}
\setcounter{equation}{0}

One can hardly overestimate the importance of supergravity
solutions. The solutions describing the long-range fields
associated with strings, D-branes and solitonic fivebranes
have played an instrumental role in many advances in string
theory. String dualities require the existence of certain
solutions and conversely the pattern of supergravity solutions
strongly hints of similar patterns and properties of the
underlying microscopic theory. Furthermore, the interplay
between the microscopic and the supergravity description of
an object has been extremely fruitful. One of the most
prominent examples is the case of black holes and their
study in string theory. One can construct solutions
describing black holes by superimposing (intersecting)
``elementary'' branes, i.e. fundamental strings, D-branes, etc.
These objects have a well-defined description in string
perturbation theory and, provided appropriate conditions hold,
one can use this description to obtain results about black holes.
For instance, such considerations led to a microscopic understanding
of the black hole entropy for extremal black holes.
Furthermore, such reasoning applied to D3 and other branes led to
the AdS/CFT correspondence, and generalizations thereof.

In all these studies, the $p$-brane solutions solve the field equations
that follow from supergravity actions that involve up to two-derivative
terms.
These actions are the lowest order terms in the
low-energy effective theories of string theories,
and the latter are known to
receive string corrections.
The corrections appear as a series in $\a'$ and are
higher derivative terms.

Given the importance of the $p$-brane solutions, one may ask how
the solutions are modified by the higher dimensional terms.
Any such modification will represent the leading stringy effects
at low energies. It is known
that some solutions do not receive any corrections.
Examples of such solutions are maximally supersymmetric spacetimes such
as flat space, and the $AdS_5 \times S^5$ vacuum of IIB supergravity,
but also spacetimes with less supersymmetry such as pp-wave
solutions \cite{KR}. These cases, however, are rather
exceptional and generically one expects the solutions to
receive corrections, see for example \cite{CMP}.
$\a'$ corrections to the near-horizon geometry of extremal
and non-extremal D3 branes were studied in \cite{BG,KR,GKT,PT}.
It was found that the $AdS_5 \times S^5$ geometry is not
corrected, but the non-extremal version is. 
Higher derivative corrections to near-horizon-NS5/little string theory
thermodynamics have been considered in 
\cite{Harmark:2000hw,Berkooz:2000mz}.

The precise form of the corrections may have implications in all
problems involving $p$-brane solutions. For instance, the $\a'$
corrections to $p$-brane solutions will induce $\a'$ corrections
to black hole solutions and their properties, such as their
entropy formula. The explanation of such subleading terms in
terms of a microscopic theory will then pose a new challenge
to our understanding of black holes.
In the context of the AdS/CFT correspondence,
$\a'$ corrections are associated with subleading terms in the
't Hooft coupling expansion. Other applications involve the
computation of $\a'$-corrections to duality transformation rules.
The higher derivative terms also become important near spacetime
singularities  where curvatures are large.

To compute the precise form of the corrected D-brane solutions, one would
need the complete set of bosonic terms in the low-energy effective action
at leading order. Higher derivative interactions can be computed by
scattering  amplitudes \cite{GW,GS} or using sigma model techniques
\cite{Grisaru:vi} (see \cite{paper} for a more
complete list of references).  However, apart from the well known $R^4$ term
only a very few other terms are known, see
\cite{Frolov:2001xr,Peeters:2001ub,FT}
for recent discussions. One way to obtain further interactions
terms is to find all terms that are related by supersymmetry to
the known terms.
In \cite{paper} we investigated in detail the possibility
of constructing a superinvariant as a scalar superpotential term in
IIB superspace \cite{HW}. A linearized version of such
a term was known to contain the $R^4$ term \cite{Green1}
leading to the expectation that such a superspace term
contains all terms that are related to the $R^4$ by linearized supersymmetry.
We have shown in \cite{paper}, however, that a superinvariant based on a
scalar
superpotential does not exist.
Finding the superinvariant associated to the $R^4$ term,
and thus determining  the complete set of interactions at leading order
is still an open question. For the computation of the corrections to
D-brane solutions one
would need the full set of bosonic terms depending on the metric,
dilaton and RR-fields.

In this paper we systematically analyze the computation
of corrections to brane solutions. The computation consists of
obtaining the corrected field equations,
evaluating the terms that originate from the higher derivative terms on
the lowest order solution and then integrating the resulting equations.
We present several different methods to obtain the field
equations. The straightforward determination of the field
equations is possible but very laborious. A method that we
find well-suited for this problem is the Palatini formalism.
In this formalism the metric and the Christoffel symbols
are considered as independent fields that are varied
independently. The simplifications are due to the fact that one
has to perform fewer partial integrations
when deriving the field equations. This reduces the number of
terms that participate in the field equations. This formulation,
even though simpler than the direct computation, is still tedious.

A significant improvement is possible when one considers
spherically symmetric solutions. In this case we derive
an effective one-dimensional action that governs
the field equations. This action may be thought
of as the consistent reduction of the ten dimensional action
over all coordinates but the radial one. The method developed
can also be used more generally to
derive consistent reductions in general. The field equations
to be solved are second order differential equations. In the
case where the lowest order solution is supersymmetric, we also
derive associated first order equations that include the effects
of the higher derivative terms (we present such an analysis for D3
branes, but similar considerations are applicable to other
branes as well).

After the field equations are derived, we have to evaluate the
higher derivative terms on the lowest order solution whose
corrections we want to compute. This leads to $r$-dependent
source terms in the field equations. To explicitly
compute the source terms one needs to know the exact form
of the higher derivative terms which is not known at present.
Given such source terms, however, we succeeded
in integrating the equations to obtain the corrections
as integrals of the sources. When the higher derivative terms
become available, these results would immediately lead to
the exact form of the corrected solutions.

One of the cases that is under better control is the case
of the D3 brane. In this case the lowest order solution
has a constant dilaton and a self-dual five-form.
This eliminates some of the possible interaction terms.
For instance, higher derivative terms that depend on the derivatives
of the dilaton will not contribute and thus they need not be considered.
Even in the D3-brane case, however,
there are possible yet undetermined interaction terms
depending on the 5-form RR field $F_5$ and derivatives thereof
(the superpotential term mentioned above does contain such terms).
In fact, our analysis indicates that such terms will contribute to
the full form of the corrected D3-brane solutions.
Noting this, we proceed by taking into account the corrections due to
the $R^4$ term only. In this sense, the computation may be viewed as
a toy model computation. 
We obtained the corrected solution in closed form.
It has a non-trivial dilaton, 
is regular in the interior and approaches $AdS_5 \times
S^5$ in the near-horizon limit.

In the presence of higher derivative interactions the standard
formulae for the computation of the thermodynamic properties of the
solutions are modified. We discuss in detail,
following \cite{Wald1,Wald2,Wald3}, how to do such computations.
We find that the tension and the electric
potential of the D3 brane renormalize, but the charge, temperature and
entropy remain uncorrected. Despite the renormalization
of the tension, we show that a BPS-type formula
that relates the mass and the charge still holds.
This formula follows from the integrated form of the first
law of thermodynamics (Smarr formula).
The renormalization of the mass is
compensated by the renormalization of the electric potential.

Any correction to the mass of the D3 brane due to higher derivative
terms has rather dramatic consequences: the mass
of $N_1$ branes plus the mass of $N_2$ branes is higher than
the mass of $N_1+N_2$ branes. This implies that there is a force
between the branes and the branes will tend to coalesce together.
This is opposite to what expects from BPS branes. We take these
results as a strong indication that the higher derivative
terms contain $F_5$ dependent terms so that there are additional
contributions to our computation.

One may expect that once the $F_5$ terms are included, the full
extremal D3-brane solution will turn out to be uncorrected, but such a
proof is still lacking. Such non-renormalization will be consistent with
the fact that KK-monopole solution, which is connected to the
D3 brane via dualities, does not receive corrections from the
$R^4$ term. This follows from the fact that the corresponding
sigma model is finite \cite{NonR}.
(Since the KK-monopole is a purely gravitational
solution there is no issue of undetermined higher derivative
interactions). This argument, however, assumes that the duality
rules will not introduce any $\a'$ corrections, but in general
the T-duality rules are known to receive
$\a'$ corrections, see for instance \cite{Kaloper:1997ux}.
Another way to analyze this question would be to study
Killing spinor equations but the corrections to supersymmetry
rules due to the higher derivative corrections are also not yet
available.

This paper is organized as follows. In the first three sections
we analyze in detail the corrections to the D3 brane due to the
$R^4$ term. In particular, in section 2 we discuss the
derivation of the corrected field equations. We present three methods:
the direct derivation of the field equations, the application of the Palatini
method and the derivation of an effective one-dimensional
action. The analysis in this section holds for both extremal
and non-extremal branes (but some of the explicit formulas apply only to
extremal D3 branes). In section \ref{fo} we restrict our attention
to the extremal D3 brane and rewrite the equations of motion in first
order form, which we then integrate to obtain the $\a'$ corrected
solution. In section 4 we discuss in detail
thermodynamics for higher derivative theories and apply the results
to the corrected D3 brane solution.
In section \ref{otherbranes} we discuss the corrections to
extremal electric $p$-branes in $D$ dimensional spacetimes.
We conclude with a discussion of our results in section \ref{disc}. Finally in
appendix A and B we give several results regarding the
the evaluation of the higher derivatives terms on lowest
order solutions, and in appendix C we present the most general
D3 brane solution of the lowest order equations with a specific
two-block ansatz.

\section{Equations of motion} \label{eqm}
\setcounter{equation}{0}

The fields that participate in the D3-brane solution of IIB supergravity
are the metric $g_{ij}$, the dilaton $\f$, and the
four-form gauge field $A_{i_1...i_4}$.
The terms in the classical IIB supergravity action that only involve
these fields, in the Einstein frame, read\footnote{
Our curvature conventions are
$R_{ijk}{}^l=\pa_j \G_{ik}^l - \G^l_{ip} \G^p_{jk} - (i \leftrightarrow j)$,
$R_{ij}=R_{ikj}{}^k, R=g^{ij} R_{ij}$. The Weyl tensor is
given by $C_{ijk}{}^l = R_{ijk}{}^l
+ {1 \over 8} [\d^l_i R_{jk} + g_{jk} R^l_i - {1 \over 9} R \d^l_i g_{jk}
-(i  \leftrightarrow j)]$.},
\be \label{iib}
I=-{1 \over 16 \p G_N} \int\dd^{10}x \sqrt{-g} [R - \half (\pa \f)^2
-{g_s^2 \over 4 \cdot 5!} F_5^2]
\ee
where
\footnote{Notice that we use the convention \label{conv}
of leaving a factor of $g_s$ in Newton's
constant. This means that our ``Einstein frame'' is related to
the string frame by $\tilde{g}_{E}=e^{-(\f-\f_\infty)/2} g_{st}
=g_s^{1/2} g_E$, where $g_E$ is the true Einstein metric and
$e^{\f_\infty}=g_s$. Under S-duality $g_E$ is invariant,
but $\tilde{g}_E \to g_s^{-1} \tilde{g}_E$.}
$G_N=8 \p^6 g_s^2 \a'^4$. The field equations derived
from this action should be supplemented by the self-duality (SD) condition
on $F_5$.

The leading higher derivative terms in the low energy effective
action of IIB string theory appear at order $\a'^3$.
The purely gravitational terms can be computed by the 4-point
graviton scattering amplitudes \cite{GW} or a four-loop sigma
model computation \cite{Grisaru:vi} and give rise to the well-known
$R^4$ terms.
To compute the corrections to the D3-brane solution we need to know
the higher derivative terms that involve $g_{ij}$, $\f$ and
$F_5$.\footnote{We assume throughtout this work that the 
fields that are zero in the lowest order solution remain zero 
after the $\a'$ corrections are taken into account. This 
would be correct if all higher derivative terms are at least 
quadratic in the fields that are zero at lowest order.}
As discussed in the introduction,  the complete set
of such terms is not known  at present.
In principle, such terms can be computed by studying
tree-level scattering amplitudes. One would need to compute up
to 8-point functions in order to compute all 8-derivative terms in the
effective action. Terms that depend on the RR-fields are more difficult
to compute using the sigma model methods in the RNS formalism,
but one could use sigma models in the pure spinor formalism \cite{pusp,BH}
to perform a manifestly supersymmetric beta function computation,
see  \cite{dBS} for such a computation.

We will proceed by considering only the effect of the $R^4$ term.
This is what has been done in similar computations in most of the
literature. We emphasize, however, that there is no
a priori reason that the $\a'{}^3$ terms can be truncated to only the $R^4$
term. In fact our results indicate that, at least for the computation of
$\a'$ corrections to the D3-brane solution, the truncation is not
consistent. We consider the following $\a'{}^3$ corrections to (\ref{iib}),
\be \label{Sw}
I_{W}=-\frac{1}{16 \p G_N} \int\dd^{10}x \sqrt{-g}\, 
\g(\f)
W
\ee
where 
\be \label{Wdef}
\g(\f)={1 \over 16} E_{3/2}(\f)g_s^{3/2} \a'^3, \quad
W = C^{imnj} C_{kmnl} C_i{}^{rsk} C^{l}{}_{rsj} +
\frac{1}{2} C^{ijmn} C_{klmn} C_i{}^{rsk} C^{l}{}_{rsj}~,
\ee
Notice that we used the field redefinition ambiguity
\cite{Tseytlin,GW} to reach a scheme where $W$ depends only on the
Weyl tensor. $E_{3/2}(\t,\bar{\t})$ is the
non-holomorphic modular form
of weight (0,0)\footnote{Explicitly,
$E_{3/2}(\t,\bar{\t}) = \sum_{(m,n) \neq (0,0)} 
{\t_2^{3/2} \over |m+n \t|^3}$, where $(m,n)$ denotes the 
greatest common divisor of the integers $m$ and $n$.
A non-holomorphic form $F^{(w,\hat{w})}$ of weight $(w,\hat{w})$
transforms as
$F^{(w,\hat{w})} \to F^{(w,\hat{w})} (c \t + d)^w (c \bar{\t} + d)^{\hat{w}}$
under the $SL(2,Z)$ transformation $\t \to (a \t + b)/(c \t + d)$.}.
Here $\t=\t_1 + i \t_2 = \chi + i e^{-\f}$, where $\chi$ is the axion. 
In the following we set $\chi=0$. The factor 
of $g_s^{3/2}$ in (\ref{Wdef}) is correlated with our conventions, 
see footnote \ref{conv}. The dilaton dependence follows from 
supersymmetry and the $SL(2,Z)$ symmetry of IIB string theory 
\cite{Green1,E32}. This behavior takes into account non-perturbative
effects as well. At string tree-level $\g(\f)|_{tree} = \frac{1}{8} 
\zeta(3)$.

The equations of motion of IIB supergravity in the Einstein frame,
restricted on these fields, and including the
corrections from (\ref{Sw}), read,
\bea \label{eqn}
E_{ij} &\equiv&
R_{ij}- \half \,g_{ij}\,R -{1 \over 2}
[\pa_i\f\pa_j\f -\half\,g_{ij}\,(\pa\f)^2]
-{g_s^2 \over 96}\,(F_{il_1\ldots l_4} F_j{}^{l_1\ldots l_4}
-{1 \over 10} g_{ij} F_5^2) \nonu
&&+\,(w_{ij}-\half\,g_{ij}\, \g(\f) \,W)=0 \\
E &\equiv& \Box\f
-\g_\f(\f)
W =0 \label{feqn} \\
F_5&=&\star F_5 \label{sdeqn}
\eea
where\footnote{ 
$D_w=i (\t_2 \pa/\pa \t - iw/2)$ and $
\bar{D}_{\hat{w}}=-i (\t_2 \pa/\pa \t + i\hat{w}/2)$ are 
modular covariant derivatives that map a modular form of weight 
$(w,\hat{w})$ to another one of a different weight,
$D_w F^{(w,\hat{w})} = F^{(w+1,\hat{w}-1)}$, 
$\bar{D}_{\hat{w}} F^{(w,\hat{w})} = F^{(w-1,\hat{w}+1)}$.}
\be 
\g_\f = 
{\pa \g \over \pa \f} = - {1 \over 32} \a'^3 g_s^{3/2}(D_0 + \bar{D}_0) E_{3/2}
\ee
and 
$w_{ij}$ is defined by
\be \label{wij}
\int d^{10} x \sqrt{g}\, \g(\f)  \d W =
\int d^{10} x \sqrt{g}\, \d g^{ij} w_{ij}
\ee
and is given in Appendix \ref{direct}. Using the fact that the Weyl tensor is Weyl invariant
one can show that
\be
g^{ij} w_{ij} = 4\, \g(\f) W~.
\ee

Notice that the self-duality equation (\ref{sdeqn}) is expected to
receive corrections from the $\a'{}^3$ terms that depend on $F_5$.
The reason is the following. An $F_5$-dependent $\a'$ correction will
give rise, upon variation w.r.t. the gauge field,
to the equation
\be \label{sdmo}
\frac{1}{ \sqrt{-g}} \frac{1}{2 \cdot 4!}
\pa_l ( \sqrt{-g} F^{li_1..i_{5}}) +
\g\, w_A{}^{i_1..i_{5}} = 0
\ee
where $w_A{}^{i_1..i_{5}}(g,A,\phi)$ is the variation of the extra
term w.r.t. the gauge field. Suppose now that the self-duality
condition holds. The first term in (\ref{sdmo}) would then vanish
by itself and we obtain,
\be
w_A{}^{i_1..i_{5}} (g,A,\phi) = 0
\ee
We thus find a new equation arising at order $\a'^3$.\footnote{
This conclusion would be avoided if there is a higher
derivative term that depends on $F_5$ and is not zero
on-shell (w.r.t. the lowest order equations), but whose variation
vanishes on-shell. As far as we can tell, this cannot happen.}
The higher derivative terms, however, should only correct the lowest
order equations, not introduce new equations. Any new equations
would generically make the system of equations inconsistent.
It follows that if the higher derivative terms are $F_5$ dependent,
the self-duality condition will have to be deformed. In other 
words, it should be a combination of the 5-form field strength 
with other fields that is self-dual not the 5-form by itself.
Notice that any superinvariant
based on the dilaton superfield will contain  $F_5$ dependent
terms \cite{paper}.
Since in this work we only take into account (\ref{Sw}) the self-duality
equation holds at order $\a'{}^3$ as well.
This is one point where the complete analysis is expected to deviate
from the analysis presented here.

We look for perturbative solutions in $\a'$
of these equations of motion. The general ansatz we consider is
\be \label{ansatz}
ds^2 = e^a \left((- f dt^2  + d \vec{x}^2)
+ e ^h (f^{-1} dr^2 + r^2 d\O_5^2)\right)
\ee
where the functions $a$, $h$ and $f$ depend only on the radius $r$.
Extremal solutions have $f=1$, but
we will keep $f$ arbitrary for the time being, and set $f=1$
at a later stage.  The self-duality condition
is solved by,
\be \label{sdsol}
F_{tabcr}  =  16 \p N \a'^2 \epsilon_{abc} e^{-2 h} r^{-5}, \qquad
F_{m_1...m_5} = 16 \p N \a'^2 \e_{m_1...m_5}
\ee
where $a,b,c$ are spatial worldvolume coordinates, $m_1,...m_5$ are indices
on the $S^5$ directions and $\e_{abc}$ and $\e_{m_1...m_5}$
are the volume densities on flat $R^3$ and on the unit five-sphere, $S^5$,
respectively.

The lowest order equations of motion admit the solution,
\be \label{Lorder}
e^{-2 a_0} = e^{h_0}=1 + {\ell^4 \over r^4}, \qquad e^{\f_0}=g_s, \qquad
f=1, \qquad \ell^4 = 4 \p g_s N \a'^2
\ee
where  the subscript in $a_0, h_0$, and $\f_0$  indicates that this
is the lowest order solution. The solution
describes the long range field of $N$ D3-branes. Removing
the ``1'' from the harmonic function yields
$AdS_5 \times S^5$, the near-horizon limit of the D3-branes.

Our objective is to obtain a solution of the equations
of motion (\ref{eqn}) perturbatively in $\a'$, i.e. we will look for
solutions
\be \label{peran}
a=a_0 + \g a_1, \qquad
h=h_0 + \g h_1, \qquad
\f=\f_0 + \g_\f \f_1, \qquad f=1,
\ee
To obtain $a_1$, $h_1$ and $\f_1$
one may substitute the ansatz (\ref{ansatz}) with the coefficients
in (\ref{peran}) to the field equations (\ref{eqn}) and keep only
the terms of order $\a'^3$. The computation involves evaluating the
order $\a'^3$ terms in (\ref{eqn}) on the lowest order
solution (\ref{Lorder}). We now present a few different formulations
of the problem.

\subsection{Direct computation} \label{dir}

This is the straightforward approach where one first obtains
$w_{ij}$ by varying the new term in the $10d$ action and then substitutes
the
lowest order solution. Both of these steps are straightforward but
very tedious. The general expression for $w_{ij}$ is given in
appendix \ref{direct}. The evaluation of the corrections on
the lowest order solution is also very tedious
because the expressions involve tensors with complicated index
contractions. A useful observation is that one can use the
symmetries of the Weyl tensor to rewrite (\ref{Wdef}) in the
following compact form
\be
W = B_{ijkl} (2 B^{iklj} - B^{lijk})
\ee
where
\be
B_{ijkl} = C^m{}_{ijn} C^n{}_{lkm}
\ee
This tensor is symmetric under a pair interchange and under simultaneous
permutation of the first two and last two indices,
\be
B_{ijkl} = B_{klij}, \qquad B_{ijkl}=B_{jilk}.
\ee
The use of Mathematica was instrumental
in obtaining the final equations. We will present these equations after
presenting two alternative methods for performing the computation.

\subsection{Palatini formalism}

There is an alternative method to derive field equations that is
particularly useful in higher derivative actions. We outline it here because
it is completely general and can be used when there are no symmetries that
one
can employ to derive a simple form of the action (as we do in the next
subsection). Furthermore, this method is simpler than the direct
derivation of the equations of motion described in the previous
subsection. In this method one constructs a Palatini action
that is first rather than second order in derivatives, and has the metric
and the
covariant derivative (or, equivalently, the Christoffel symbols)
as independent variables (see \cite{Wald} for an elementary exposition):
\be
I[g,\G]=\int\dd^{10}x\,\@{g}\left(R[g,\G] -\half(\pa\phi)^2
-{g_s^2\over4\cdot5!}\,F_5^2 +\g(\f) \, 
W[g,\G]\right).
\ee
In deriving the equations of motion, let us vary the connection first. This
gives:
\bea
\d^{(j}_i \na_l g^{k)l} - \na_i g^{jk} + (g^{jk} \d^p_i - \d^{(j}_i g^{k)p})
(\G^l_{lp} - \pa_p \log \sqrt{g}) + 
W_i^{jk} =0~,
\eea
which at lowest order in $\a'$ implies the usual compatibility condition
between
the metric and the connection. Standard manipulations yield:
\be
\G^k_{0ij}=\half\,g_0^{kl}(\pa_ig_{0jl} +\pa_jg_{0ik} -\pa_lg_{0ij})~.
\ee
We also find the following solution at next order:
\bea
\G^k_{1ij}&=& -{1\over18} (\d^k_iW_{jl}^{\,\,\,\,\,\,l}
+\d^k_jW_{il}^{\,\,\,\,\,\,l}) +\half \,(W_{j\,\,\,\,i}^{\,\,\,k}+
W_{i\,\,\,\,j}^{\,\,\,k} - W^k_{\,\,\,ij})\nn
\G^j_{1ij}&=&-{1\over 9}\,W_{ij}^{\,\,\,\,\,\,j},\hspace{1cm}
g^{jk}\G^i_{1jk}=-{1\over 9}\,W^{ij}_{\,\,\,\,\,\,j}.
\eea
The r.h.s. of the above formulas should be read as being evaluated in the
lowest
order metric. $W_i{}^{jk}$ is such that
\be
\int 
\g(\f)\,
\d W = \int 
\d g^{ij} W_{ij} +\int \d \G^i_{jk} 
W_i{}^{jk}.
\ee
These tensors satisfy the following identities:
\bea
g^{ij}W_{ij}&=&4 \g(\f) W\nn
W^j{}_{ji}&=&\half W_{ij}{}^j.
\eea
By explicit computation one finds that $W_i{}^{jk}$ is given by the covariant
derivative of a tensor that is cubic in the Weyl tensor, although we will
not
give the explicit expression here. Symbolically, $W_i{}^{jk}$ has the
structure:
$W\sim \na(\g(\f)CB)$ where $B$ is the tensor defined in the previous section.
At the
end of the day, the combination that appears in the equation of motion is
given
in terms of a {\it single} scalar function of $r$ when computed
for the lowest
order solution, but we will not report the details here.

The remaining equations for the metric and matter fields are also easily
derived. In particular, deriving the equation of motion for the metric is
much
simpler than in the second order formalism. One obtains equations of motion
where the Ricci tensor depends on both $g$ and $\G$. One then expands this
in
the above solutions to obtain the standard form of the Einstein equations:
\bea
R_{ij}-\half\,\pa_i\f\pa_j\f -{g_s^2\over96}\,
F_5^2{}_{ij}
+ 
[W_{ij}-{3\over8}\, g_{ij} \g(\f)\, W] &+&\nn
+\half\,\na_k[W_{j\,\,\,\,i}^{\,\,\,k} + W_{i\,\,\,\,j}^{\,\,\,k} -
W^k_{\,\,\,ij}]&=&0\nn
\Box\f 
-\g_\f \,
W&=&0,
\le{palatinieom}
which are supplemented with the self-duality condition. These equations are
equivalent to the ones found by direct computation, but their
derivation is simplified.

\subsection{Effective $1d$ action}

We show in this subsection that for spherically symmetric solutions, there
is an effective one-dimensional action that yields the
same field equations as (\ref{eqn}) evaluated on the ansatz (\ref{ansatz}).
To obtain the one-dimensional action we start from the variation of the
ten-dimensional action,
\be \label{var1}
\d I = \int d^{10}x\,\sqrt{g}\, [\d g^{ij} E_{ij} + \d \f E]~.
\ee
where we have substituted the solution (\ref{sdsol}) of
the self-duality equation (\ref{sdeqn}) in $E$ and $E_{ij}$.
We now use the ansatz (\ref{ansatz})
to express $\d g^{ij}$  in terms of $\d a, \d h,$ and $\d f$.
This yields
\be \label{var2}
\d I = \int d^{10}x\, \sqrt{g}\, [- \d a  (g^{ij} E_{ij})
+{\d f \over f} (g^{rr} E_{rr} - g^{tt} E_{tt})
-\d h (g^{rr} E_{rr} + g^{mn} E_{mn})
+ \d \f E]~.
\ee
Since all the fields depend only on the radial variable, one can
now perform all integrations but the radial one.
The resulting variations can be integrated again to an one dimensional
action,
\bea
I_{1d}&=& \int dr e^{4a + 2 h} \frac{r^5}{\ell^5} \left[
  \frac{1}{6}
\left( 64 f a''+  40 f h'' + 148 f a'^2 +
       168  f a' h' + 50 f h'^2 + 3  f \phi'^2  \right. \right. \nonu
&& \left. \qquad
       + 400 \frac{f a'}{r} +  240 \frac{f h'}{r} + 200 \frac{f}{r^2} +
       6 f'' + 64  f' a'  + 37 f' h'
       + 80  \frac{f'}{r} - \frac{120}{r^2} \right) \nonu
&& \left. \qquad
+\frac{8 \ell^8}{r^{10} e^{4(a+h)}} - \g(\f) e^{a+h} W
\right] \label{1daction}
\eea
where we have discarded an overall (infinite) volume factor.
$W$ is given by (\ref{Wdef}) evaluated on (\ref{ansatz}). It is a function
of   $a$, $h$, $f$ and their first two radial derivatives. The
explicit expression is given in appendix A.

Notice that this derivation of the effective action guarantees
that all solutions of the $1d$ action are solutions of the
$10d$ action. In other words, the reduction from
$10d$ to $1d$ is consistent. What is crucial is that the number of
independent functions appearing in the ansatz (\ref{ansatz})
is equal to the number of equations one gets by evaluating
(\ref{eqn}) on the (\ref{ansatz}). For the problem at hand
this number is four even when $f=1$, so even in this case
one must first proceed with general $f$ and then set $f=1$.
The method presented here can be used more generally in
order to provide consistent reductions of the higher
dimensional theories. One should contrast this method
with the most common practice to substitute an ansatz in the
action and then reduce. This latter does not guarantee a consistent
reduction.

It is instructive to rewrite (\ref{1daction}) in terms of
the variables used in the reduction of the type IIB supergravity over $S^5$.
Such a reduction was presented
in \cite{GKT}. Using their variables the one-dimensional action reads
\bea \label{5daction}
I_{1d} &=& -\int dr \sqrt{g_5} \left[R_5 - \frac{1}{2}
g_5^{rr}(\partial_r \phi)^2
-\frac{40}{3} g_5^{rr} (\partial_r \nu)^2 - V(\nu)
+ \g(\f) e^{-\frac{10}{3} \nu} W \right] \\
V(\nu) &=&
\frac{1}{\ell^2}
\left[8 e^{-\frac{40}{3} \nu} -
20 e^{-\frac{16}{3} \nu} \right] \label{potential}
\eea
where $g_5$ denotes the determinant of the five dimensional
metric $g_{5mn}$ given in (\ref{5dmetric}).
The fields appearing in this action are related to the $a, h$ and $f$ by
\bea
\nu(r) &=& \frac{1}{2} (a+h) + \log{\frac{r}{\ell}} \\
ds_5^2 =g_{5 mn} dx^m dx^n &=& e^{ {1 \over 3}(8 a + 5 h)}
\left({r \over \ell}\right)^{{10 \over 3}}
(-f dt^2 + \d_{ab} dx^a dx^b +
\frac{e^h}{f} dr^2) \label{5dmetric}
\eea
This can be shown using the standard reduction formula
\be
ds_{10}^2 =e^{-\frac{10}{3} \nu} g_{5 mn} dx^m dx^n +
e^{2 \nu} \ell^2 d \Omega_5^2
\ee
and matching with the ansatz in (\ref{ansatz}). The dimensionful
parameter $l$ is proportional to the Planck length and is
introduced into the ansatz on dimensional grounds.

The equations of motion that follow from (\ref{1daction}) with $f$ set equal
to one, $f=1$, are given by
\bea
&&18 a'' + 10 h''+ 36 a'^2 + 36 a'h' + 10 h'^2 + \phi'^2
+ \frac{90 a'}{r} + \frac{50 h'}{r}
+ \gamma w_a =0 \label{eaex} \\
&&10 (a'' + \half h'') + 2 a'^2 + 20(a'+\half h')^2 +
\frac{1}{2} \phi'^2 + \frac{50}{r} (a' + \half h')
- \frac{8 \ell^8}{r^{10} e^{4(a+h)}}   \nonu
&& \qquad + \gamma w_h =0 \label{ehex} \qquad \\
&&8 a'' + 5 h''  - 4 a'^2 - 4 a' h' + \phi'^2+ \frac{5 h'}{r}
+\gamma w_f =0  \label{efex} \\
&&\phi''  + (4 a' + 2 h' + \frac{5}{r} )\phi'
- 2 \gamma_\f w_{\f} =0 \label{ephiex}
\eea
where $\g w_a$ is the variation of $\a'^3$ term in action
(\ref{1daction}) with respect to $a$, etc.\ . We give
in appendix \ref{direct} the explicit form of $W$ as a function
of $a$, $h$, $f$ and their derivatives. From there one may
derive $w_a$ etc. The evaluation of the corrections
on the lowest order solution  (\ref{Lorder})
is given by
\bea \label{ws}
w_a &=& w_\f =  - 14400
\frac{\ell^{16}}{r^{24} e^{{19 \over 2} h_0}} \nonu
w_h  &=& -\frac{4800 \ell^{12}}{r^{28} e^{{19 \over 2} h_0}}
(112 \ell^8-249 \ell^4 r^4 + 84 r^8)  \\
w_f &=& \frac{28800 \ell^{12}}{r^{28} e^{{19 \over 2} h_0}}
(14 \ell^8- 35 \ell^4 r^4 + 10 r^8)  \nonumber
\eea
Notice that the metric in (\ref{ansatz}) depends on $a$ only through
an overall conformal factor. It follows that
the Weyl tensor does not depend on $a$, and that $W$
depends on it only through the inverse metrics used in contracting
indices. This explains why $w_a$ is equal to $w_\f$.

We have explicitly verified that the equations (\ref{eaex})-(\ref{ephiex})
are equivalent to the equations one obtains by evaluating (\ref{eqn}) on the
ansatz (\ref{ansatz}), as discussed in subsection \ref{dir}.
This remains true even when $f$ is not set equal
to one. This a nice check, especially on
$w_a, w_h, w_f$ and $w_\phi$, as the organization of the two
computations is rather different.
In the next section we present
yet another reformulation in terms of first order equations.

\section{First order system} \label{fo}
\setcounter{equation}{0}

The D3 brane solution is half supersymmetric \cite{DL1}.
This implies that there must be an equivalent first order formulation
of the field equations when the ansatz for the solution is consistent
with supersymmetry.
In this section we set $f=1$, and present such a reformulation.
A (somewhat different) discussion of first order equations
appeared in \cite{FT}.

Let us first consider the effective action without the $\a'$ correction.
The potential in (\ref{potential}) has an AdS critical point at
$\nu=0$.  This critical point is stable
as it is maximally supersymmetric. It follows that the potential
$V(\nu)$ admits a ``superpotential'' $\cw$ such that the AdS critical
point is a critical point of $\cw$ \cite{Town}. Indeed, one finds that
the potential (\ref{potential}) can be rewritten as
\bea \label{potential1}
V(\nu) &=& {3 \over 10}
\left(\frac{\pa \cw_{\pm}}{\pa \nu} \right)^2 - \frac{16}{3} \cw_{\pm}^2 \\
\cw_{\pm} &=&
\frac{1}{\ell}
\left[e^{-\frac{20}{3} \nu} \pm \frac{5}{2}
e^{-\frac{8}{3} \nu} \right]~.
\eea
The formula for the potential (\ref{potential1}) coincides with the one
in \cite{Town} after
the differences in conventions are taken into account.  The AdS
critical point is also a critical point of $\cw_-$.
We shall henceforth consider only $\cw_-$, which we shall denote $\cw$,
and only add a few comments about $\cw_+$.

A simple Bogomolnyi argument implies that the theory
admits BPS domain wall solutions \cite{ST,DFGK}
\bea \label{DW}
ds_5^2 &=& e^{2 c(\r)} \eta_{ab} dx^a dx^b + d\r^2 \\
\nu&=&\nu(\r) \nonumber \\
\f &=&\f(\r) \nonumber
\eea
where $c(\r)$ and $\nu(\r)$ are solutions of the first order equations
\be \label{1st}
\pa_\r \nu = {3 \over 20} \frac{\pa \cw}{\pa \nu}, \qquad
\pa_\r c = -\frac{2}{3} \cw, \qquad
\pa_\r \f=0~.
\ee
One can verify that solutions of the first order system solve the second
order equations.
The first order equations also follow from the requirement that the
``Killing
spinor'' equations
\be \label{suru}
(D_\m + {1 \over \sqrt{15}} \cw \G_\m) \e =0, \qquad
(\G^\m \pa_\m \nu  - {3 \over 5 \sqrt{2}} {\pa \cw \over \pa \n}) \e =0,
\qquad
\G^\m \pa_\m \f \e =0
\ee
admit solutions for non-zero spinor $\e$ \cite{ST}. In the context of
supergravity
these are the variations of the gravitino and dilatino, and
the solutions of the first order equations are supersymmetric solutions.

The coordinate transformation,
\be \label{tr}
r=r(\r), \qquad {d \r \over d r} =
\left({r \over l}\right)^{5/3} e^{{4 \over 3} (a(r) + h(r))},
\ee
can bring the metric (\ref{5dmetric}) to the form (\ref{DW}).
Furthermore, $a(r)$ and $h(r)$ are related to $c(\r)$ and
$\nu(\r)$ in a simple way,
\bea \label{nuc}
\nu(\r)&=&\frac{1}{2} [a(r(\r))+h(r(\r))] + \log {r(\r) \over l}\\
c(\r) &=&
{1 \over 3} [4 a(r(\r))) + {5 \over 2} h(r(\r)) + 5 \log
{r(\r) \over l}] \label{nuc1}
\eea
It follows that one can obtain first order equations for
$a$ and $h$ by substituting  (\ref{nuc}) and (\ref{nuc1})
in (\ref{1st}).  One obtains,
\be \label{susy}
\pa_r a + \pa_r h + {2 l^4 \over r^5} e^{-2 (a+h)} =0, \qquad
4 \pa_r a + {5 \over 2} \pa_r h + {2 l^4 \over r^5} e^{-2 (a+h)} =0, \qquad
\qquad \pa_r \f=0.
\ee
These are exactly the equations that follow from
the analysis of supersymmetry in ten dimensions \cite{DL1}.

Before we move on to consider the modification due to $\a'$ corrections
we note that had we considered the superpotential $\cw_+$, we
would have ended up with a solution of the form (\ref{Lorder})
but with
\be
e^{h_0} = 1 - {l^4 \over r^4}, \quad r^4 > l^4, \qquad
e^{h_0} = -1 + {l^4 \over r^4}, \quad r^4 < l^4
\ee
This solution has a curvature singularity at $r=l$, and
is related to the standard D3 brane solution by analytic
continuation to imaginary $r$.

\subsection{$\a'$ corrections to the first order system}

We now discuss the extension of the analysis to include the $\a'$
corrections.
Ideally one would like to write the effective action as a sum and/or
differences of
squares and then read off the $\a'$-corrected first order equations.
Such a rewriting should be possible because of supersymmetry.
However, the complexity of $W$ for general $h, a$ and $\f$
makes such an exercise rather formidable. Furthermore, as
we discussed, our action is not complete since further relevant
bosonic terms may be present and such additional terms may be
necessary in order to rewrite the action as a sum of squares.

We proceed by adding order $\a'^3$ terms in (\ref{susy}) and demand that
the solutions of the first order system solve the second order equations
(\ref{eaex})-(\ref{ephiex}),
\bea \label{1stal}
a' + h' + {2 l^4 \over r^5} e^{-2 (a+h)} &=& \gamma j_1(r) \\
4 a' + {5 \over 2} h' + {2 l^4 \over r^5} e^{-2 (a+h)} &=& \gamma 
j_2(r)
\label{1stal2} \\
\f' &=&2 \c_\f  j_3(r)  \label{aord}
\eea
where prime indicates a derivative w.r.t. $r$.  This yields
\bea
j_1&=&2(1+{10 \over r h_0'}) b_1 - {1 \over 2 h_0'} (w_h + w_f - w_a), \nonu
j_2&=&5(1+{4 \over r h_0'}) b_1 - {1 \over 2 h_0'} (w_h + w_f - w_a), \nonu
j_3 &=& \frac{1}{r^{5}}\int^r dr' r'^{5} w_\f + {C_1 \over r^{5}} \label{j3}
\eea
where $b_1 \equiv a_1'+\half h_1'$ satisfies
\be \label{bode}
b_1'+{9 \over r} b_1 = {1 \over 10} ( w_f - w_a)
\ee
Notice that supersymmetry demands that $b_0=0$
to lowest order.
There are non-supersymmetric solutions of the lowest
order second order equations (\ref{eaex})-(\ref{ephiex}),
including non-supersymmetric solutions with $b_0=0$,
as we discuss in the appendix \ref{gene}, but we shall not consider them
here.

Once  $w_f$, $w_a$ and $w_h$ are computed using the lowest order solution,
(\ref{bode}) for $b_1$ and the equation for $j_3$ can be easily integrated.
$b_1$ in turn gives the source terms $j_1$, and $j_2$.
The integration constants are fixed by requiring that
solution is asymptotically flat and
regular at the horizon.

\subsection{$\a'$-corrected solution}

Using the $w$'s in (\ref{ws}) one can easily compute the
sources,
\bea
j_1(r) &=&
C_0 \frac{ 3 \ell^4 + 5 r^4}{r^{9} \ell^4}
- \frac{16 \ell^{12}}{2431 r^{43} e^{\frac{17}{2} h_0} } \left[
768 \ell^{24} + 7808 \ell^{20} r^4
+35360 \ell^{16} r^8 \right.  \nonu
&&
\left. + 93 840 \ell^{12} r^{12} +
161330 \ell^8 r^{16} +
365 86 55 \ell^4 r^{20} - 3500640 r^{24} \right] \\
j_2(r) &=&
\frac{5 C_0}{r^5 \ell^4}
- \frac{320 \ell^{12}}{2431 r^{39} e^{\frac{17}{2} h_0}}
\left[ 64 \ell^{20} + 544 \ell^{16} r^4
+2040 \ell^{12} r^8
+ 4420  \ell^8 r^{12}
\right. \nonu
&& \left.
+ 133705  \ell^{4} r^{16} -
109395  r^{20} \right] \nonu
j_3(r)&=&
\frac{C_1}{r^5} - \frac{160 \ell^{16}}{
     2431 r^{39}  e^{\frac{17}{2} h_0}}
     \left( 128\,\ell^{16} + 1088\,\ell^{12}\,r^4 +
       4080 \,\ell^8\,r^8  + 8840\,\ell^4\,r^{12} +
       12155\, r^{16} \right). \nonumber
\eea
The integration constants $C_0$ and $C_1$ can be fixed
by requiring that the terms on the right hand side
of (\ref{1stal})-(\ref{aord}) are small
compared to the terms on the left hand side for all $r$.
This implies that $j_1, j_2$ and $j_3$ should be
at most the same order as the terms on the left hand side.
Near $r=0$ the terms on the left hand side behave as
$1/r$. On the other hand, $j_1$ behaves as $1/r^9$ and
$j_2$ and $j_3$ as $1/r^4$. This can be remedied by choosing
appropriately the integration constants,
\be \label{intcon}
C_0={2^{12} l^2 \over 2431}, \qquad C_1 = {5 \over l^4} C_0~.
\ee
This is a non-trivial result since the number of terms that
we need to set to zero is greater than the integration
constants we have.
The same values of integration constants follow by requiring
that the solution we present in the next section is smooth at the horizon.

We note that the $j_1$ and $j_2$ are such that they cannot be absorbed into a
$\a'$-modification of the superpotential $\cw$. To check this one
may rewrite (\ref{1stal}) in the coordinate system (\ref{tr}).
Let us call $J_1(\r)$ and $J_2(\r)$ the sources that appear on the
right hand side of the first and second equations in (\ref{1st}).
One may absorb $J_2(\r)$ into $\cw$ by  $\cw'= \cw -{3 \over 2} \g J_2$.
In order for this transformation to also remove the source $J_1$ the following
relation should hold,
\be
{\pa J_2 \over \pa \r} + {2 \over 3} {\pa \nu \over \pa \r} J_1 =0~.
\ee
A direct computation shows that this is not satisfied, but we note that
there are (unexpected) cancellations between the two terms. Had we
been able to absorb the sources into a modified superpotential,
we would conclude that the form of the supersymmetry rules
in (\ref{suru}) is not modified at order $\a'^3$, so these
results may be taken to indicate that there are new terms in the
supersymmetry
transformation rules at order $\a'^3$. We should add however that
given that we only consider a part of the complete effective action
such a conclusion is premature.

Knowing the sources,
it is straightforward to integrate the first order equations
(\ref{1stal})-(\ref{aord}).
Taking the sum of the first two equations one obtains a differential equation
for $a+h/2$ which can be easily integrated. Feeding back one solves
for $a_1$ and $h_1$. The integration of (\ref{aord}) is
equally straightforward. All integration constants are set to zero
by requiring that the solution is asymptotically flat. The result is
\bea
h_1 &=& - \frac{1024 (3 \ell^8 + 9 \ell^4 r^4 + 10 r^8)}{
2431 \ell^2 r^{12} e^{h_0}}
- \frac{32 \ell^{12}}{2431 r^{38} e^{{15 \over 2} h_0}}
\left[ -96 \ell^{20}  - 912 \ell^{16} r^4 - 3910 \ell^{12}
r^8  \right. \nonu
&-& \left. 22 355  \ell^8 r^{12} - 97240 \ell^4 r^{16} +
218 790 r^{20} \right] \nonu
a_1 &=& \frac{1024 (2 \ell^8  + 5 \ell^4 r^4 + 5
r^8)}{2431 \ell^2 r^{12} e^{h_0}}
+ \frac{8 \ell^{12}}{2431 r^{38} e^{{15 \over 2} h_0} } \left[
-256 \ell^{20} - 2304 \ell^{16} r^4  \right.
\nonu
&-& \left. 9384 \ell^{12} r^8 -47600 \ell^8 r^{12}
- 197795 \ell^4 r^{16} +
486 200 r^{20} \right] \nonu
\phi_1 &=&
-\frac{10240}{2431 l^2 r^4}
+ \frac{160 \ell^{16}}{2431 r^{34} e^{{15 \over 2}h_0}}
\left[ 64 \ell^{12} + 408 \ell^8 r^4 + 1020 \ell^4 r^8
+ 1105 r^{12} \right] \label{corrsol}
\eea
The corrections are smooth at $r=0$, and the choice of integration
constants was crucial for this property.

Let us consider the near horizon limit of the solution.
Following \cite{Malda} we consider the limit,
\be \label{near}
\a' \to 0, \qquad {r \over \a'}\ {\rm fixed}, \qquad g_s  \ {\rm fixed}
\ee
In this limit we find that
\be
\g h_1 = - \g a_1
={1 \over N^{3/2}} {E_{3/2}(g_s) \over 2 \cdot 2431 \p^{3/2}} , \qquad
\g_\f \f_1 = - 180 {\g_\f \over \g} (\g h_1)
\ee
It is intriguing that even though we ignored higher derivative
terms that depend on $F_5$ the near horizon limit is still
$AdS_5 \times S^5$, just as one would expect for the ``true''
D3 brane solution \cite{BG,KR}. This may indicate that $F_5$
dependent terms can be ignored in the near-horizon limit.
Recall that in the near-horizon limit $F_5$ is proportional
to the volume form both in the $AdS_5$ and the $S^5$ directions.
One may verify using the results in \cite{paper}
that the $F_5$ dependent terms of the dilaton superfield
vanish in this case. This is an additional indication
that our results are exact in this case.

Notice that the AdS radius does not receive corrections
but the string coupling constant does. The choice of the
integration constants in (\ref{intcon}) is crucial for
the limit (\ref{near}) to exist.

\section{Thermodynamics of corrected solutions} \label{sol}
\setcounter{equation}{0}

In this section we discuss the thermodynamics of the
corrected solution.  The quantities of interest are the
mass density\footnote{Notice that
we use interchangeably the terminology ``mass density''
and ``tension''. With abuse of terminology will also sometimes
just call ``mass'' the tension, and ``charge'' the charge
density. It will be clear, however, from the context
which quantity we are discussing.}, the temperature, the entropy and the
charge density of the solution.  One may use either Euclidean
of Lorentzian methods to study thermodynamics. In the
present case the self-duality of the lowest order solution
presents an additional complication in the Euclidean
computation since one needs to understand the proper
analytic continuation of the self-duality condition.
We will follow a Lorentzian analysis
and adapt the method of Wald \cite{Wald1,Wald2,Wald3}
for the problem at hand.

Recall that the entropy, mass and charge of a black hole
satisfy the first law of thermodynamics which in
integrated form (Smarr formula) reads
\be \label{smarr}
T S = M - v Q
\ee
Here $T$ is the Hawking temperature, $S$ is the entropy, $M$ is the
mass, $Q$ is the charge and $v$ the corresponding potential.
Extremal black holes have zero temperature $T=0$ (and quite often zero
entropy as well) so the Smarr formula implies,
\be \label{masscharge}
M=v Q
\ee
In the context of supersymmetric black holes, this relation
originates from the supersymmetry algebra. The case of
D-branes is exactly analogous, but the appropriate quantities
are now densities. One may wrap the spatial worldvolume
coordinates of the brane on a torus (or some other compact manifold)
and reduce over that manifold to obtain a black hole
in lower dimensions. For instance, the D3 brane can be viewed
as a $7d$ black hole after reduction over the spatial
worldvolume coordinates.
Our analysis will be done from the ten dimensional point of view.

In the presence of higher derivative terms, the extremal D3 brane
still has zero temperature (as we verify below), so a relation of the
form (\ref{masscharge}) should still hold since (\ref{smarr}) follows
from first law alone. Since the charge of the D3 brane is quantized one
might expect that (\ref{masscharge}) would imply that the mass does
not renormalize. We find, however, that things are more subtle
and both the mass and the potential $v$ renormalize.

Given a gravitational system described by an action $I$ one
may compute the gravitational energy as follows.
Let us consider a spacetime $M$ and denote by
$\pa M_\infty$ its asymptotic infinity which is
considered as its boundary.
We first require that the theory,  subject to appropriate
boundary conditions, has a well-defined variational
problem, i.e.
all boundary terms in the variation of the action should vanish
automatically so that the bulk field equations are true
extrema of the action. In gravitational systems this requires the
addition of boundary terms $B$,
\be \label{action}
I = \int_M L - \int_{\pa M_\infty} B.
\ee
Under a variation we have
\be
\delta L = (\mbox{{\rm field\ equations}}) + d \Theta(\F,\delta \F)
\ee
where $\F$ denote collectively all fields. In order for the variational
problem to be well-defined
$B$ and $\Theta$ should be related by
\be
\d \int_{\pa M_\infty} B = \int_{\pa M_\infty} \Theta(\F,\delta \F)
\ee
In pure gravity $B=2 K$,
where $K$ is the trace of the second fundamental form.
In more general theories $B$ may contain additional terms.

The action (\ref{action}) is invariant under diffeomorphisms.
This implies that there is a
corresponding Noether current,
\be
{\bf J} = {\bf \Theta}(\F, \cl_\xi \F) - i_\x {\bf L}
\ee
where $\xi^a$ is a vector that generates the diffeomorphism,
$\cl_\xi$ is the Lie derivative along $\xi^a$,
$i_\xi$ is the inner derivative (when acting on
a $n$-form it produces a $(n-1)$-form by contracting its
first index by $\xi^a$) and we use form
notation (${\bf J}$ is a $(d-1)$-form, ${\bf L}$ is a $d$-form etc.).
When the field equations are satisfied
${\bf J}$  is closed, $d {\bf J}=0$, and one can  construct locally
a $(d-2)$-form ${\bf Q}$ such that ${\bf J}=d {\bf Q}$.
The Hamiltonian that provides the dynamics generated by $\xi^a$ is
given by
\be
H = \int_C {\bf J} - \int_{\pa M_\infty} i_\xi {\bf B}
\ee
where $C$ is a Cauchy surface. On-shell this evaluates to a
surface term
\be \label{energy}
H =  \int_{\pa M_\infty}({\bf Q} - i_\xi {\bf B})
\ee
The gravitational energy is now defined by taking $\xi$ to be a
timelike Killing vector. In general, this expression is
divergent so a suitable subtraction should be employed.
In asymptotically AdS spacetimes one may incorporate
in ${\bf B}$ covariant boundary counterterms \cite{count},
but in asymptotically flat spacetimes such universal
covariant local counterterms do not exist \cite{countfl}.
We (implicitly) use the background subtraction method below.

Let us now consider the theory based on the action (\ref{iib}).
Following similar steps as in \cite{Wald2} one finds
that the mass density of a D3 brane solution
is given by
\be \label{ADM}
\mu={M \over V}
= \frac{1}{16 \p G_N} \int \left ( \pa_m h_{pm} - \pa_p h^j_j \right) dS^p_5
\ee
where $h_{ij}$ is the deviation of the metric from the Minkowski
metric and $V$ is the volume of the spatial worldvolume directions.
The integration is over the sphere at asymptotic infinity in transverse
space. The index $j$ runs over all spatial indices and $p$ and $m$ only
over the  transverse coordinates.
Formula (\ref{ADM}) generalizes the ADM formula to apply to
$p$-brane spacetimes \cite{DGHR}. One may rewrite this formula as
a Komar-like mass formula,
\be \label{komar}
M={1\over8\p G_N}\int_{S_\infty}\e_{a_1\ldots a_8bc}\nabla^{[b}\xi^{c]}
\ee
where $S_\infty$ is the spacelike surface at infinity enclosing the brane.
We
wrap the spatial worldvolume coordinates of the brane on a torus of volume
$v$, so that $S_\infty=T^3\times S^5$. Static spacetimes of the form
(\ref{ansatz}) have a a timelike Killing vector
\be
\x=\x^i {\pa \over \pa x^i}, \qquad \xi^t =1, \qquad \x^i = 0, i \neq t
\ee
A straightforward computation yields
\be \label{xir}
\nabla^r \x^t = \half g^{rr} g^{tt} \pa_r g_{tt}.
\ee
One may now substitute this expression in (\ref{komar})
to obtain the mass of the solution.

The temperature $T$ associated with a spacetime is equal to $T=\k/2 \pi$
where $\k$ is the surface gravity. The latter can be show to be
equal to \cite{Wald}
\be \label{temp}
\k^2 = - \half \nabla^a \xi^b \nabla_a \xi_b
= -{1 \over 4} g^{tt} g^{rr} g_{tt,r}^2
\ee
where in the last step we used (\ref{xir}).

The entropy of a solution can be computed using the $(d-2)$-form
${\bf Q}$ introduced earlier,
\be \label{ent}
S = \int_{\ch} {\bf Q} = -\pi\int_{\ch} \dd\S_{ij}\,Q^{ij}
\ee
where
$\dd\S_{ij}=\dd^8x\sqrt{h}\,\e_{ij}$ is the surface element defined over the
horizon $\ch$, with $h$ the induced metric on the horizon. $Q_{ij}$ is
related
to the Noether charge as discussed above and (after fixing ambiguities
with certain choices) is given by
\be \label{entropy}
Q^{ij}=-2{\cal L}^{ijkl}\na_k\xi_l +4\na_k{\cal L}^{ijkl}\xi_l.
\ee
${\cal L}^{ijkl}$ is the variation of the action with respect to the Riemann
tensor and $\xi$ is the timelike Killing vector.
In the case where the action contains only the Einstein-Hilbert term, the
result given the well-known Bekenstein-Hawking formula, $S=A/4$.
The derivation of (\ref{entropy}) assumes a non-degenerate horizon
($\k \neq 0$). It was successfully applied, however, (in a context
similar to ours) to extremal black holes as well \cite{deWit}.\footnote{
We thank Bernard de Wit for discussions about this point.}
We will assume that this formula remains valid for extremal black
holes.

Finally the electric charge density of the solution is given by
\be \label{charge}
q = \frac{g_s}{\sqrt{16 \pi G_N}} \int_{S^5} *F_5
\ee
The prefactor is due to the normalization of the
$F_5$ terms in (\ref{iib}). The magnetic charge $\tilde{q}$ is given by
a similar integral that involves $F_5$. In general, the electric
and magnetic charges satisfy the Dirac quantization condition
\cite{dirac}
\be \label{dcon}
q \tilde{q} = 2 \pi n
\ee
where $n$ is an integer. For dyons this formula is modified
and it does not by itself lead to a quantization condition
for self-dual solutions with $q=\tilde{q}$, such
as the D3 brane solution which we discuss \cite{DL1}.
The exact quantization condition for D3 branes is
determined in string theory by string dualities (see for
instance section 3 of \cite{bhrev}).
With the normalizations as in (\ref{Lorder})
the charge $q$ of a single self-dual brane comes out to be $q=\sqrt{2 \pi}$
(see (\ref{chd3})), which agrees with the naive application of (\ref{dcon})
with $n=1, q=\tilde{q}$.

\subsection{Lowest order solution}

Before we proceed to incorporate the $\a'$ corrections let us discuss
the lowest order D3 brane solution. In this case the metric is
given in (\ref{Lorder}), and the mass density can be
easily calculated (using either (\ref{ADM}) or (\ref{komar})) to be
\be
\mu = {N \over (2 \pi)^3 g_s \a'^2}
\ee
where we used $G_N= 8 \pi^6 g_s^2 \a'^4$. The charge density of the solution
is given by
\be \label{chd3}
q = \sqrt{2 \pi} N,
\ee
where the factor of $\sqrt{2 \pi}$ is discussed below (\ref{dcon}).
It is straightforward to use the formulas given above to
compute the entropy and the temperature of the solution. The result is that
both of them are equal to zero. So one expects a formula of the form
(\ref{masscharge}) and we indeed find
\be \label{bps}
\mu = {1 \over \sqrt{16 \pi G_N}}\, q~.
\ee
This is the BPS formula derived in \cite{DL1}. Let us now derive this
relation in a way that will be useful when we consider the corrected
solution.

Using Stokes' theorem one can express the surface integral in
(\ref{komar}) in terms of a volume integral,
\be \label{sto}
M = - {1 \over 4 \p G_N} \int_{\S} \e_{a_1...a_9 b} R^b_c \x^c
+{1 \over 8 \p G} \int_{\ch} \e_{a_1...a_8 b c} \nabla^{[b} \x^{c]}
\ee
where $\S$ is a spacelike hypersurface that extends from the
horizon to spatial infinity, and
the last term is a surface integral over the horizon (which also
involves the worldvolume $T^3$).
To derive this one needs to use
\be
\nabla_j \nabla^j \x^i = - R^i_j \x^j
\ee
which holds for Killing vectors.
The integral over the horizon may be evaluated using our
explicit metric and it vanishes. In general, this term gives
the entropy contribution in the first law.

To evaluate the volume term we now use the Einstein equation,
\be \label{einst}
R_{ij} = \half \pa_i \f \pa_j \f
+ {g_s^2 \over 96} F_{il_1\ldots l_4} F_j{}^{l_1\ldots l_4}
\ee
We further note that $\x$ generates an isometry of the solution, so
\be \label{inv}
\x^i \nabla_i \f=0, \qquad
\x^i \nabla_i A^{j_1...j_4} + 4 (\nabla^{[j_1} \x_k) A^{|k| j_2 j_3 j_4]}=0
\ee
Inserting (\ref{einst}) in (\ref{sto}) we get a term that depends on the
dilaton and a term that depends on $F$. The former yields a vanishing
contribution upon using (\ref{inv}). The latter can be manipulated as
follows,
\bea \label{id}
\xi^i F_{il_1...l_4} F^{jl_1...l_4} &=&
- 4 \nabla_{l_1} (\x^i A_{i l_2 l_3 l_4} F^{jl_1 l_2 l_3l_4})
+(\x^i \nabla_i A_{l_1 l_2 l_3l_4}
+ 4 (\nabla_{[l_1} \x^k) A_{|k| l_2 l_3 l_4]})
F^{jl_1 l_2 l_3 l_4}  \nonu
&&-4 \x^i A_{i l_2 l_3 l_4} \nabla_{l_1} F^{jl_1...l_4}
\eea
The last two terms vanish due to the $F$-field equation and the invariance
of the solution (\ref{inv}). We finally get
\be
M = -{g_s^2 \over 96 \p G_N} \int_{S_\infty \cup \ch} \e_{tra_1...a_8}
\xi^i A_{i l_1 l_2 l_3} F^{trl_1 l_2 l_3}
\ee
One may integrate $F_{t123r}$ to obtain $A_{t123r}$,
\be
A_{t123} = {1 \over g_s} (e^{-h_0}-1)
\ee
where the constant part was chosen such that $A_{t123}$
vanishes asymptotically. It follows then that
$\xi^i A_{i 123}|_\ch = -1/g_s$. In general, one can change the
asymptotic value of $A_{t123}$ by performing a gauge transformation.
This will modify the value of $\xi^i A_{i 123}$ at the horizon.
The combination
\be \label{lowv}
v=\xi^i A_{i 123}|_{S_\infty}-\xi^i A_{i 123}|_\ch = - {1 \over g_s}
\ee
is the associated electric potential and is gauge invariant.\footnote{On a
curved spacetime one may define the
electric and magnetic part of a a field strength as
\be
{\bf E} = i_\xi {\bf F}, \qquad {\bf B} = i_\xi *{\bf F}
\ee
where ${\bf E}$ and ${\bf B}$ are four-forms in our case,
$i_\xi$ is the inner derivative and $\xi$ is a timelike Killing vector.
For self-dual solutions, ${\bf E}={\bf B}$.
When the field equations and Bianchi identity hold,
$d {\bf F} = d * {\bf F}=0$,
one finds that $d {\bf E} = d {\bf B} =0$, so locally there are
electric and magnetic potentials ${\bf E} = d {\bf v}$ ,
${\bf B} = d \tilde {\bf v}$, respectively. \
In the case at hand, the electric potential is related with the gauge field
as ${\bf v}=i_\xi {\bf A}$.
One may show in general that ${\bf v}$ is constant at the horizon
and the difference between its asymptotic value and the constant value
at the horizon is gauge invariant.}
It is this gauge invariant combination
that couples to the electric charge $q$ (notice that one
may use Stokes' theorem to show that $\int_\ch *F = \int_{S_{\infty}} *F$).
It follows that
\be
\mu = {1 \over \sqrt{16 \p G_N}}\, (- g_s v)\, q
\ee
which is equal to (\ref{bps}) upon using (\ref{lowv}).

In the present case we were able to explicitly manipulate
the bulk integral in (\ref{sto}) into boundary terms.
When we include the $\a'$ corrections, however,
similar manipulations involving the higher derivative
term become increasing complicated. Instead of using
(\ref{id}) in order to manipulate the bulk integral
one could also just use the explicit solution to evaluate the
bulk integral. Since the dilaton is constant the first term
on the right hand side drops out. The contribution of
$F_5$ can also be computed straightforwardly since the integral
can be computed by elementary means. These manipulations
lead to the same result (\ref{bps}), but now the contribution
of the charges was computed via a bulk integral.

\subsection{Corrected solution}

The $\a'$ corrected D3-brane solution in the Einstein
frame is given by
\bea \label{solution}
&&ds^2 = e^{-\half h_0}\left(1+\g a_1\right)(- dt^2  + d \vec{x}^2)
+ e^{\half h_0}
\left(1+ \g (a_1 + h_1)\right) (dr^2 + r^2 d\O_5^2) \nonu
&&e^\f = g_s(1 + \g_\f \f_1) \\
&&F_{tabcr}  = 16 \p N \a'^2 \epsilon_{abc} e^{-2 h_0}
(1-2\g h_1)
r^{-5}, \qquad
F_{m_1...m_5} = 16 \p N \a'^2 \e_{m_1...m_5} \nonumber
\eea
where the $e^{h_0}$ is given in (\ref{Lorder}) and $a_1, h_1$ and
$\f_1$ are given in (\ref{corrsol}). We should emphasise that
this solution would be the true corrected D3 brane only if the
part of the effective action relevant for this problem
consisted of only (\ref{iib}) and (\ref{Sw}). However, as we
discussed earlier, it is likely that additional terms
that dependent on $F_5$ are relevant.

To compute the mass of the solution one has to take into
account that the action has been modified by the addition
of the term (\ref{Sw}), so the mass formula should also
be modified accordingly. The discussion in the beginning
of this section outlines the steps that are necessary
in order to compute the new mass formula. This computation
is technically rather complex because of the complicated
tensor contractions in $W$. We will use instead the following
shortcut. We will start from (\ref{komar}) and use Stokes'
theorem to rewrite it as in (\ref{sto}). We should emphasise
that the starting point does not represent the entire mass
of the corrected solution since it does not properly take
into account that the mass formula has been modified.
We now use the field equations,
\be \label{einst1}
R_{ij} = \half \pa_i \f \pa_j \f
+ {g_s^2 \over 96} F_{il_1\ldots l_4} F_j{}^{l_1\ldots l_4}
+ \left({3 \over 8} \g(\f) W g_{ij} - w_{ij} \right)
\ee
The first two contributions can be analyzed as in the previous
subsection. The last contribution represents an additional
gravitational contribution. It should thus be combined with
the term on the left hand side to yield the mass of the solution.
We thus propose as a mass formula
\be \label{massform}
M = {1\over8\p G_N}\int_{S_\infty}\e_{a_1\ldots a_8bc}\nabla^{[b}\xi^{c]}
+{1 \over 4 \pi G_N} \int_\S \e_{d a_1 \cdots a_9}
\left({3 \over 8} \g(g_s) W g^{d}{}_e - w^{d}_{e} \right) \xi^e
\ee
The logic here  is similar to the one discussed in the
last paragraph of the previous subsection: one could either
rewrite the bulk integral as a surface integral or just
directly compute the bulk integral.

The result for the mass is
\be   \label{masscor}
\mu=\mu_0
\left(1 + \g(g_s) {5 \cdot 2^{10} \over 2431} {1 \over l^6}\right)
=\m_0
\left(1 + {1 \over N^{3/2}} {40 E_{3/2}(g_s) 
\over 2431 \pi^{3/2}} \right)
\ee
where $\mu_0= N/(2 \p)^3 g_s \a'^2$ is the mass density of the lowest
order solution. 
In this result the three terms in (\ref{massform}) contribute to
the correction with relative weights $-1, 3/2, 0$.

The form of the correction in (\ref{masscor})
follows by dimensional analysis and the
fact that the lowest order solution depends on the parameters of the
solution only via $l^4$. The detailed form of the higher derivative
term only determines the numerical coefficient. In particular, if the
numerical coefficient is non-zero, as we find in (\ref{masscor}),
then the mass of $N_1+N_2$ branes is less than the mass of
$N_1$ branes plus the mass of $N_2$ branes. This follows from
the inequality
\be
{1 \over \sqrt{N_1 + N_2}} <
{1 \over \sqrt{N_1}}+{1 \over \sqrt{N_2}}.
\ee
It follows that energetically the branes would prefer to coalesce to form a
single group. Thus there should be a force acting on the
two sets of branes. This is opposite to what one expects
from BPS configurations, where the branes should not feel
any force. We believe that after taking into account the
effect of the (presently unknown) $F_5$ dependent higher derivative
terms the mass of the D3 brane solution will not renormalize.

With the definition of mass in (\ref{massform}) one may
proceed as in the previous section to derive
\be
M = -{g_s^2 \over 96 \p G_N} \int_{S_\infty \cup \ch} \e_{tra_1...a_8}
\xi^i A_{i l_1 l_2 l_3} F^{trl_1 l_2 l_3}
\ee
One may integrate $F_{t123r}$ to obtain $A_{t123}$,
\bea
A_{t123}&=&{1 \over g_s} (e^{-h_0}-1)
-{32 \pi N \g(g_s) \a'^2 \over 2431 l^2 r^{40} e^{9 h_0}}
\left( 256 r^{28} e^{7 h_0} (3 l^4 + 5 r^4) \right.  \nonu
&&+ 16 l^{14} r^2 e^{\half h_0}
(- 48 l^{16} - 352 l^{12} r^4 - 1037 l^8 r^8 + 1700 l^4 r^{12} + 19890
r^{16}))
\eea
and from here we obtain
\be
v=\xi^i A_{i 123}|_{S_\infty}-\xi^i A_{i 123}|_\ch = v_0
(1 + \g(g_s) {5 \cdot 2^{10} \over 2431}{1 \over l^6})
\ee
where $v_0=-1/g_s$ is the value of the electric potential
for the lowest order solution. The remaining of the
computation is exactly the same as the one in the previous
paragraph, and we end up with
\be \label{smarr2}
\mu = {1 \over \sqrt{16 \p G_N}}\, (- g_s v)\, q
\ee
The charge density of the solution retains its
lowest order value, as is required by charge quantization.
We thus find that even though the mass of the solution renormalizes
and the charge does not renormalize, a BPS-type formula
still holds. This is possible because the electric potential renormalizes.

One can understand this behavior
as follows. In the absence of corrections to the gauge field
equation, the formula for the charge,
$q \sim \int * d A$, remains uncorrected. Since $q$ does not
renormalize and $*$ renormalizes (since it depends on the metric),
$A$ has to renormalize in such a way that the combined
corrections to $*$ and $A$ cancel each other. So unless the gauge field
equation is corrected the electric potential will renormalize.
Then the first law (\ref{smarr2}) can be used to infer that the mass
renormalizes. As we argued above, however, any correction
to the mass would imply that the branes feel a force.
This strongly indicates that the gauge field equations,
and therefore the self-duality condition of $F_5$
receives corrections such that at the end the mass
of the brane does not renormalize.

One may easily check that the temperature and the entropy
remain equal to zero. For the temperature this follows
upon using (\ref{temp}). It goes to zero as $r$, as in lowest
order solution, but the coefficient of $r$ receives corrections.
For the entropy we use (\ref{ent})-(\ref{entropy}).
The corrections to the entropy vanish as $r^{15}$ (after factoring out
the behavior of the temperature).

\section{Other branes} \label{otherbranes}
\setcounter{equation}{0}

Corrections to other R and NS $D$-brane solutions can be analyzed
as in the D3-brane case. Analogously to in the 3-brane case, an
$1d$ dimensional  effective action may be derived. The system of second
order equations  can be integrated by introducing
variables suggested by the lowest order supersymmetry relations. As in the
D3 case, these equations contain source terms evaluated on the lowest
order solution. Once the complete source terms are known,
the corrected solutions can be derived.

\subsection{Equations of motion}

The $D$ dimensional action in the Einstein frame relevant for
general brane solutions is given by
\bea
S=-\frac{1}{16 \pi G} \int\dd^Dx \sqrt{-g} \left( R- {4\over D-2}(\pa\phi)^2
- {1\over2(p+2)!} e^{\tilde{a} \phi} F_{p+2}^2
+ \g e^{-{12\over D-2}\phi} W \right)
\eea
where
\be
\g = \frac{1}{8} \zeta(3) \alpha'^3 g_s^{{12\over D-2}} \qquad
\tilde{a} = \frac{2(D-2 p -4)}{D-2} - a_{NS}
\ee
$a_{NS}=2$ for NS branes but zero otherwise. The dilaton factor in front
of $W$ is that of a tree-level string correction. $W$ is expected to depend
on $F_{p+2}$ and on its covariant derivatives as well as on the covariant
derivatives of the dilaton. As discussed in the introduction, this
expression
is not known at present, so in our analysis we will keep $W$ arbitrary.

The equations of motion from the above action are
\bea
E_{ij} &=& R_{ij} - \frac{1}{2} g_{ij} R - \frac{1}{2}
\left( \pa_i \phi \pa_j \phi
- \frac{1}{2} g_{ij} (\pa \phi)^2 \right) \\
&& \hspace{-1cm}-\frac{e^{\ta \phi}}{2(p+2)!} \left(
(p+2) F_{il_1 \ldots l_{p+1}} F_{j}^{l_1 \ldots l_{p+1}} - \frac{1}{2}
g_{ij} F^2_{p+2} \right)
+ \gamma \left( w_{ij} -
\frac{1}{2} g_{ij}\,e^{-\frac{12}{D-2} \phi}  W \right) =0 \nonu
E&=& \na^2 \phi- \frac{\ta e^{\ta \phi}}{2(p+2)!} F^2_{p+2} +
\g w_\f =0
\\
E^{i_1 \ldots i_{p+1} } &=& \frac{1}{ \sqrt{-g}} \frac{1}{(p+1)!}
\pa_l ( \sqrt{-g} e^{\ta \phi} F^{li_1..i_{p+1}}) +
\g w_A{}^{i_1..i_{p+1}} = 0
\eea
where $w_{ij}$ is defined by
\be
\int\dd^Dx\sqrt{g}e^{-{12\over D-2}\f}\d W=\int\dd^Dx\sqrt{g} \d
g^{ij}w_{ij}
\ee
and $w_\f$ and $w_A$
denote the variation of $e^{-\frac{12}{D-2} \f} W$ w.r.t. the dilaton
and the gauge field, respectively.

The equations of motion admit both electric and magnetic
solutions \cite{HS,DL2,DKL}. Here we will consider
only electric branes, but similar methods apply
to magnetic branes as well.
For an electric brane solutions we take the ansatz
\bea \label{ansatze}
ds^2 &=& e^a \left( -dt^2 + \sum_{a=1}^{d-1} (dx^a)^2   + e^h (dr^2 +
r^2 d \Omega_{q+1}^2 )\right) \nonu
A_{t\a_1...\a_p} &=& \e_{\a_1...\a_p} c(r).
\eea
The field equations evaluated on this ansatz give
\bea
&&a'' + \frac{(D-2)}{2} a'^2 + \frac{q}{2} a' h' + (q+1)\frac{a'}{r}  -
e^{\tilde{a} \phi} \frac{q}{D-2} K^2 + \gamma \tilde{w}_t =0 \label{e1} \\
&& (D-1) a'' + (q+1) h'' - \frac{d}{2} a' h' + \phi'^2 + \frac{(q+1)}{r} a'
+  \frac{(q+1)}{r} h' - e^{\tilde{a} \phi} \frac{q}{D-2} K^2 \nonu
&& \qquad+ \gamma
\tilde{w}_r =0 \label{e2}  \\
&& a'' + h'' + \frac{(D-2)}{2} a'^2 + \frac{1}{2} (2q + d) a' h' +
\frac{1}{2} q h'^2  + (2q + d + 1) \frac{a'}{r} + (2q + 1) \frac{h'}{r}
\nonu
&& \qquad + e^{\ta \phi} \frac{d}{D-2} K^2
+ \gamma \tilde{w}_m =0 \label{e3} \\
&& \phi'' + \left(\frac{(D-2)}{2} a' + \frac{q}{2} h'+ \frac{(q+1)}{r}
\right) \phi' + \frac{\ta}{2} e^{\ta \phi} K^2 + \g \tilde{w}_{\phi} =0
\label{e4} \\
&&{1 \over (p+1)!} e^{-{D \over 2} a - (q+2) h - \ta \f} r^{-(q+1)}
\pa_r \left(e^{({D \over 2} -p-2) a + \half q h +\ta \f} r^{(q+1)} k(r)
\right)
+ \g \tilde w_A =0 \label{geq}
\eea
where $d=p+1$ and  $q=D-p-3$ and
\bea
\tilde{w}_{t} &=& 2 e^{h}
\left[ \frac{1}{D-2} (-g^{kl} w_{kl} + e^{-\frac{12}{D-2} \phi}
W ) g_{tt} + w_{tt}
\right] \\
\tilde{w}_{r} &=& -2 
\left[ \frac{1}{D-2} (-g^{kl} w_{kl} + e^{-\frac{12}{D-2} \phi}
W ) g_{rr} + w_{rr}
\right] \\
\tilde{w}_{m} &=& -2 e^{a+h} 
\left[ \frac{1}{D-2} (-g^{kl} w_{kl} + e^{-\frac{12}{D-2} \phi}
W ) g_{mm}
 + w_{mm} \right] g^{mm} \\
\tilde{w}_{\phi} &=& g_{rr} w_{\phi} \\
\tilde{w}_{A} &=& e^{- \tilde{a} \phi} w_A \\
K^2 &=& k(r)^2 e^{-d a}, \qquad k = {1 \over p+2} c'
\eea
and
\be
w_A{}^{i_1 \cdots i_{p+1}} = w_A \e_{i_1 \cdots i_{p+1}}.
\ee
The same equations also hold for magnetic solutions.
One only has to exchange $q \leftrightarrow d$ and
take $\tilde a \to - \tilde a$.

The lowest order equation admits the following electric solution
\cite{HS,DL2,DKL}
\bea
e^{h_0} &=& \left[ 1 +
\left(\frac{l}{r} \right)^q \right]^{\frac{4}{\Delta}}, \qquad
a_0 = \frac{-q}{D-2} h_0, \nonu
\phi_0 &=& \frac{{\ta}}{2} h_0, \qquad
k(r) = Q e^{\left[ - \ta \phi_0 - \frac{q}{2} h_0 +
(d+1 - D/2) a_0 \right]} r^{-(q+1)}, \nonu
\Delta &=& {\ta}^2 + \frac{2 d q}{D-2}, \qquad
Q^2 = \frac{4 q^2}{\Delta} l^{2 q},
\eea
where $Q$ is the charge of the solutions.

We would like to obtain perturbative solutions
\bea
a = a_0 + \g a_1 \nonu
h = h_0 + \g h_1 \nonu
\phi = \phi_0 + \g \phi_1 \nonu
c = c_0 + \g c_1
\eea
Before we proceed to integrate the equations we will present
an $1d$ effective action from which the field equations
(\ref{e1})-(\ref{geq}) can be derived.

\subsection{Effective action}

As in the case of the D3 brane, the $1d$ effective action is
most naturally written in terms of $(d+1)$ dimensional fields.
The $(d+1)$ dimensional action may be viewed as the reduction
of the $D$-dimensional theory over the sphere $S^{q+1}$ in transverse
infinity and is given by
\bea \label{1dother}
I_{d+1}& =& \int dr \sqrt{-g_{d +1}} \left[ R_{d+1} - \frac{1}{2}
g_{d+1}^{rr} \phi'^2 - {(q+1) (D-2) \over (d-1)}
g_{d+1}^{rr} \nu'^2 \right. \nonu
&&\left. \qquad +\half g_{d+1}^{rr}
\frac{e^{\ta \phi-d a}}{f (p+2)^2} c'(r)^2
- V(\nu) \right]
\eea
where
\be
V(\nu) = - \frac{q(q+1)}{l^2} e ^{- \frac{2(D-2)}{d-1} \nu}
\ee
The metric $g_{d+1}$ and the scalar field $\nu$ are related to
$a$, $h$ and $f$ by
\bea
g_{d+1,ij} dx^i dx^j &=& e^{\alpha} (- f dt^2 + \sum_{p=1}^{d-1} dx_p^2 +
\frac{e^{h}}{f} dr^2 ) \nonu
\alpha &=& \frac{1}{d-1} \left( (D-2) a + (q+1) h + 2 (q+1)
\log{\frac{r}{l}}
\right)  \nonu
\nu &=& \frac{1}{2} \left( a + h + 2 \log{\frac{r}{l}} \right)
\eea
This can be shown  by using the standard reduction formula
\be
ds_{D}^2 = e^{-2 \frac{(q+1)}{(d-1)} \nu} g_{d+1,ij} dx^i dx^j +
e^{2 \nu} l^2 d \Omega_{q+1}^2
\ee
and matching with (\ref{ansatze}).

The $\a'$ correction can be incorporated by adding to the
action the following term
\be
I_{W} = \gamma \int dr \sqrt{-g_{d +1}} e^{-\frac{2(q+1)}{d-1} \nu
- \frac{12}{D-2} \phi} W[r,a,h,\phi,f,c].
\ee

To connect the $D$-dimensional equations with the equations of the
effective action we note that
\bea
\d g^{tt} &=& - \left(\d a + \frac{\d f}{f} \right) g^{tt} \qquad
\d g^{rr}  = \left(- (\d a + \d h) + \frac{\d f}{f} \right) g^{rr} \nonu
\d g^{mn} &=& - (\d a  + \d h) g^{mn} \qquad
\d A_{i_1 \ldots i_{p+1}} = \epsilon_{i_1 \ldots i_{p+1}} \d c
\eea
Substituting in the $D$-dimensional action we obtain
\bea \label{intout}
\delta I &=& \int d^{D} x \sqrt{-g} \left[ - E_{ij} g^{ij} \d a +
\frac{\d f}{f} \left( g^{rr} E_{rr} - g^{tt} E_{tt} \right) -
\left( g^{rr} E_{rr} + g^{mn} E_{mn} \right) \d h + E \d \phi \right. \nonu
&& \left. + E_{[j}{}^{j i_1 \ldots i_{p+1}} \epsilon_{i_1 \ldots i_{p+1}]}
\right] \nonu
&=& \int d^{D} x \sqrt{-g} \left[ E_a \d a + E_f \d f  + E_h \d h + E_c \d c
\right]
\eea
Since the solutions of interest depend on only $r$
one may integrate over all directions but $r$.
Integrating the variation to an action we get (\ref{1dother}).

{}From (\ref{intout}) we can read off the relation between the $D$
and $1d$ field equations:
\bea
E_a &=& - E_{ij} g^{ij} \nonu
E_f &=& \frac{1}{f} \left( g^{rr} E_{rr} - g^{tt} E_{tt} \right) \nonu
E_h &=& - \left( g^{rr} E_{rr} + g^{mn} E_{mn} \right) \nonu
E_{\phi} &=& E \nonu
E_c &=& E^{i_1 \ldots i_{p+1}} \epsilon_{i_1 \ldots i_{p+1}}
\eea
The sources  $\tilde{w}$ in (\ref{e1})-(\ref{geq}) can be expressed
as a combination of the source terms in the $a$, $h$, $f$, $\phi$, $c$
equations of motion:
\be
w_{a,h,f,\phi,c} = \frac{1}{\sqrt{-g}} \frac{\d}{\d(a,h,f,\phi,c)}
\left( \sqrt{-g} e^{-\frac{12}{D-2} \phi} W \right)
\ee
with the result
\bea
\tilde{w_t} &=& \frac{2 e^{(a+h)}}{d(D-2)}
\left[ q w_a - (D-2) w_h \right] \nonu
\tilde{w_r} &=& \frac{2 e^{(a+h)}}{d(D-2)} \left[ q w_a -
(D-2) (w_h + d w_f ) \right] \nonu
\tilde{w}_m &=& \frac{2 e^{(a+h)}}{d(q+1)(D-2)} \left[ (D-2)
( (1+d) w_h + d w_f )
- (q + d(q+2)) w_a  \right] \nonu
\tilde{w}_{\phi} &=& e^{(a+h)} w_{\phi} \nonu
w_{A} &=& \frac{1}{(p+1)!} w_c.
\eea

\subsection{Integrating the equations}

In this section we integrate the equations assuming that the
source terms $\tilde w$ are known. It is convenient to
introduce the following variables
\bea
b &=& a + \frac{q}{D-2} h \nonu
\Phi &=& \phi - \frac{{\ta}}{2} h
\eea
These combinations are motivated by the fact that the supersymmetry
of the lowest order solution implies that $b$ and $\Phi$ vanish
on the lowest order solution. Equations (\ref{e1})-(\ref{e4})
depend on the gauge field only through the combination $K^2$.
It is convenient to introduce the notation
\bea
\bar{K}^2 &=& e^{{\ta} \phi} K^2 \nonu
\bar{K}^2 &=& \bar{K_0}^2 + \g \bar{K_1}
\eea
where $\bar{K_0}$ is the lowest order value of $\bar{K}$
and $\bar{K_1}$ depends on the corrections.
In terms of these variable equation (\ref{e1})-(\ref{e4}) become
\bea
&& b_1'' - \frac{q}{D-2} h_1'' - \frac{q}{2} b_1' h_0' + \frac{(q+1)}{r}
b_1'
- \frac{q(q+1)}{(D-2)} \frac{h_1'}{r} - \frac{q}{D-2} \bar{K}_1 +
\tilde{w}_t =0  \label{et} \\
&& (D-1) b_1 '' + \frac{d}{D-2} h_1'' - \frac{d}{2} b_1' h_0' +
\frac{\Delta}{2} h_0' h_1' + \frac{(q+1)}{r} b_1' + \frac{(q+1)d}{(D-2)}
\frac{h_1'}{r} + {\ta} \Phi_1' h_0'   \nonu
&&- \frac{q}{D-2} \bar{K}_1 +  \tilde{w}_r =0 \label{er} \\
&&b_1'' + \frac{d}{D-2} h_1'' + \frac{d}{2} b_1' h_0' + (2 q + d + 1)
\frac{b_1'}{r} + \frac{d(q+1)}{(D-2)} \frac{h_1'}{r} + \frac{d}{(D-2)}
\bar{K}_1  +  \tilde{w}_m = 0 \label{em} \\
&&\Phi_1'' + \frac{(q+1)}{r} \Phi_1' + \frac{{\ta}}{2} \left( h_1'' +
\frac{(D-2)}{2} b_1' h_0' + \frac{(q+1)}{r} h_1' \right) + \frac{{\ta}}{2}
\bar{K}_1 +  \tilde{w}_{\phi} =0 \label{ephi}
\eea

Consider the linear combination of equations
$d (\ref{et}) + q (\ref{em})$. The resulting equation can be integrated,
\be
b_1' = \frac{1}{(D-2)}\frac{1}{r^{2 q + 1}}\left[ -\int r^{2 q + 1} \left( d
\tilde{w}_t + q \tilde{w}_m \right) + C_0 \right]
\ee
Let us introduce
\be
y = \frac{{\ta}}{2} b_1 + \frac{q}{(D-2)} \Phi_1
\ee
The linear combination  of equation $\frac{\ta{a}}{2}
(\ref{et}) + \frac{q}{(D-2)} (\ref{ephi})$ can be integrated as
\be
y' = \frac{1}{r^{q+1}} \left[
-\int r^{q+1} \left( \frac{{\ta}}{2} \tilde{w}_t + \frac{q}{(D-2)}
\tilde{w}_{\phi} \right) + Y_0 \right] \nonu
\ee
Integrating $b_1'$ and $y'$ once more one gets $\Phi_1$,
\be
\Phi_1 = \frac{(D-2)}{q} (y - \frac{{\ta}}{2} b_1)
\ee
Finally, given $b_1$ and $y$, one may integrate the linear combination
$[(\ref{er}) - (\ref{et})]$ to obtain
\be
h_1' = \frac{1}{e^{\frac{\Delta}{2} h_0} r^{q+1}} \left\{ \int r^{q+1}
e^{\frac{\Delta}{2} h_0} \left[ (\tilde w_t - \tilde w_r)
- \left( (D-2)b_1'' +
\frac{(q-d)}{2} b_1' h_0' + {\ta} \Phi_1' h_0' \right) \right] + C_3
\right\}
\ee
which can be further integrated to yield $h_1$.

Having obtained $a_1, h_1$ and $\phi_1$ one then substitutes to the
gauge field equation (\ref{geq}) to obtain $c_1$. Finally there is
one further equation among (\ref{e1})-(\ref{e4}) to satisfy
(we only used three linear combinations to obtain $a_1, h_1$ and $\phi_1$).
This is expected to follow from the other equations up to a
constant because of the Bianchi identity.
This final equation will thus impose a condition among the
integration constants, as in the case of the D3 brane.

\section{Conclusions} \label{disc}
\setcounter{equation}{0}

We have studied in this paper the computation of quantum corrections
to brane solutions. These corrections are driven by the leading
higher derivative corrections to the string theory effective
action. The corrections were computed perturbatively in $\a'$,
i.e. the lowest order solution was substituted into the
leading higher derivative term of the effective action
and then the resulting equations were integrated to obtain the
corrected solution. The straightforward application of this procedure,
i.e. the direct computation of the corrected equations,
is very tedious, basically due to the complicated tensor
structure of the higher derivative terms. We
completed the computation in this manner but we
also developed several alternative methods for analyzing the problem.

The first alternative method is the extension of the Palatini formalism
to higher derivative theories. In this method the
metric and the Christoffel symbols are considered
as independent fields. The main advantage of this
method is that it reduces the number of partial
integrations needed in order to derive the field
equations. This a significant simplification
because each factor of the Riemann tensor requires two
partial integrations in the standard derivation of the
field equations, so for higher derivative theories that
depend on $R^p$, where $R^p$ denotes $p$ Riemann
tensors contracted in some way, one gains at least
$p^2$ terms when using the Palatini method.
Furthermore, the organization of the
computation is more transparent.

Even with the simplifications of the Palatini method, however,
the computations are still very laborious. Things simplify
enormously if one is studying spherically symmetric solutions.
In this case we derived an effective one dimensional action
that governs the field equations, as we now describe.
We start by substituting in the variation of the ten dimensional action,
the ansatz for the metric  and the matter fields. Since by assumption
we are considering a spherically symmetric solution, the resulting expression
depends only on the radial coordinate $r$, and one may thus trivially
integrate over all coordinates but $r$. After discarding an overall
(infinite) volume
factor, the result can be integrated back to an one-dimensional
action where the  fields are the functions that appear in the
original ansatz. By construction the solutions of the
one-dimensional field equations automatically solve
the field equations of the original theory. Provided that
the number of independent functions in the original ansatz
is equal to the number of field equations one gets by
substituting the ansatz in the original field equations,
this method guarantees that the lower dimensional theory is a
consistent truncation of the higher dimensional one (i.e.
all solutions of the lowest order equations lift to solutions
of the higher dimensional theory).

In the cases we study in this paper, the $1d$ action has
the most transparent parametrization in terms of fields
that appear in an intermediate step. In the brane solutions
one can parametrize the transverse space in terms of polar
coordinates. The intermediate theory is obtained by
reducing over the sphere at infinity. In the context
of near-horizon geometries, this is the sphere that appears
in the near-horizon limit.
We have derived an effective $1d$ action for all D-branes
(the explicit formulas are for electric branes, but the
magnetic case can be obtained along similar lines).
These results can be used to study $\a'$
corrections to extremal and non-extremal branes, but
we only studied extremal branes in this paper.

In the case of the D3 brane we further derive first order equations.
The existence of such first order equations follows from the fact that
the potential of the intermediate theory
(obtained by the reduction over the $S^5$ at infinity)
admits an AdS critical
point (since a particular solution of our equations is the
$AdS_5 \times S^5$ solution). This implies that the lowest
equations admit a superpotential, and we indeed show
that this is the case. The inclusion of the $\a'$ corrections modifies
the first order equations by the addition of source terms.

The main limitation in our considerations comes from the
fact that the complete set of the leading higher derivative
terms is not yet available. Provided that we are supplied with
these terms, we show how to integrate the equations in all cases
to obtain the corrections to the solutions.
The corrections are given in terms of integrals of the evaluation of the
higher derivative terms on the lowest order solution.

The case of the D3 brane is under better control
because the dilaton is constant, so higher derivative
terms depending on derivatives of the dilaton
do not contribute. To compute the correction one
still needs to know the higher derivative terms that depend on $F_5$,
but these terms are not known at present. Such terms are expected
to be present because they are present in the dilaton superfield
\cite{paper}.

These terms would also lead to a modification of the self-duality
equation of $F_5$, as discussed in the introduction.
We proceed by considering the effect of only the
$R^4$ term, so
one may consider this  computation as a ``toy'' model computation.
Including the $R^4$ only, we explicitly integrate the
equations and obtain the corrected solution for this case.
We find that the integration constants may be adjusted
so that the solution is asymptotically flat and
regular in the interior. This is a non-trivial result
because the integration constants at our disposal are
less than the number of terms that are diverging in the
near-horizon limit. In turns out, however, that the coefficient
of these
terms are appropriately correlated and a smooth
limit exists. In the near-horizon limit the solution
becomes $AdS_5 \times S^5$ with the same value of the
cosmological constant but a different (constant) value
of the dilaton than the lowest order solution. The fact that
cosmological constant is uncorrected is due to
a cancellation.

We used the general method of Wald \cite{Wald1,Wald2,Wald3} to analyze the
thermodynamics
of the corrected solutions. In the presence of higher derivative terms the
ADM mass formula and other thermodynamic quantities receive corrections
and we discuss how to obtain the new formulas.
In particular, we computed all thermodynamics quantities of the
corrected D3 brane solution. We find that the temperature and the
entropy remain equal to zero, the charge is uncorrected but the
mass and the electric potential renormalize. For solutions with
zero temperature or entropy the first law of the thermodynamics
in integrated form (Smarr formula) implies that the mass density
$\mu$, charge density $q$
and electric potential $v$ are related by
\be \label{smarr1}
\mu = -{g_s \over \sqrt{16 \pi G_N}} v q
\ee
where the factor depending on Newton's constant is conventional.
The lowest order D3 brane solution satisfies (\ref{smarr1}) with
$\mu \sim q \sim N$, where $N$ is the number of D3 branes
and $v \sim g_s^{-1}$. In the corrected solution $q$ remains
uncorrected but $\mu$ and $v$ renormalize such that (\ref{smarr1})
still holds.

We emphasise again that in computing the corrections to
the D3 brane solution we did not take into account (presently
unknown) higher derivative
terms that depend on $F_5$. Such terms will
modify the field equations
(in particular the self-duality condition for $F_5$) as well as the
formula for the charges of the solution (similarly to the fact that
the $R^4$ term
modifies the ADM mass formula). A simple argument
that uses dimensional analysis and the form of the lowest order
solution shows that any (positive) correction to
the mass density would imply that it is energetically favored for
D3 branes to coalesce together rather than remain separated.
This contradicts the no-force condition
and strongly suggests that half supersymmetric
D3 brane solution do not renormalize. This in turn suggests
that the $F_5$ terms will make a significant contribution.
To properly address this issue the exact knowledge of the
$F_5$-dependent higher derivative terms will be required.

In this paper we studied corrections to extremal branes. Even though
the exact form of the higher derivative terms are not known,
we succeeded in integrating the equations in general.
Some of our
results, such the effective one-dimensional action, hold for
non-extremal branes as well.
It would be interesting to investigate whether the
non-extremal equations can be similarly integrated in general.
Other generalizations of our analysis
involve studying corrections to intersecting brane configurations.
It will be interesting to see if the simple intersection rules generalize
when $\a'$ corrections are included. This study will be relevant for
obtaining $\a'$ corrections to black hole configurations.

\section*{Acknowledgments}
KS and AS are supported by NWO.
This material is also based upon work supported by the National Science
Foundation under Grant No. PHY-9802484 and PHY-0099590.
Any opinions, findings, and conclusions or recommendations
expressed in this material are those of the authors and do not necessarily
reflect the views of the National Science Foundation.

\appendix

\section{Explicit expressions for $W$ and its variation\label{direct}}
\setcounter{equation}{0}

In this appendix we outline the details of the direct computation of the
variation of the $\a'^3$ term in the action for arbitrary $p$-branes, and present some
results for specific lowest order solutions.

We will be completely general in that we will consider a generic $p$-brane in $D$
dimensions, with an ansatz for the metric of the form:
\be\label{metricansatz}
\dd s^2=e^a\left[(-f\dd t^2+\dd\vec{x}^2_p) +e^h(f^{-1}\dd r^2+r^2\dd\O_d^2)\right],
\ee
where $a=a(r)$ and $h=h(r)$, and $p$ and $d$ satisfy $p+d=D-2$.

In a background with these symmetries the non-zero components of the Weyl tensor are:
\bea\label{xy}
C^a_{\,\,cbd}&=& Q\,(\d^a_b g_{cd}-\d^a_d g_{bc})\nn
C^t_{\,\,atb}&=& S\, g_{ab}\nn
C^r_{\,\,arb}&=& T\, g_{ab} \nn
C^m_{\,\,anb}&=& U\,\d^m_n g_{ab}\nn
C^t_{\,\,rtr}&=& V g_{rr} \nn
C^r_{\,\,mrn}&=& X\, g_{mn}\nn
C^t_{\,\,mtn}&=& Y\, g_{mn}\nn
C^p_{\,\,mqn}&=& Z\,(\d^p_q g_{mn}-\d^p_n g_{mq})~.
\eea
The explicit expressions for the functions $Q,\ldots, Z$ are:
\bea\label{x}
Q&=&-{e^{-a-h}\over4(D-1)(D-2)r^2}\left[2rf'(4d-rh'+2drh')+4(d-d^2+r^2f'')+\right.\nn
&+&\left.df\left(4drh'+ (d-1)r^2h'^2 +4(d-1+r^2h''\right)\right]\nn
S&=&-{e^{-a-h}\over4(D-1)(D-2)r^2}\left[4d-4d^2+rf'(10d-2Dd-3rh' +5drh' +Drh'-dDrh')
\right.\nn
&+&\left.6r^2f'' -2Dr^2f'' +df\left(4drh' +(d-1)r^2h'^2 +4(d-1+r^2h'')\right)\right]\nn
T&=&{e^{-a-h}\over4(D-1)(D-2)r^2}\left[-4d +4d^2+rf'(-10d +2Dd +3rh' -5drh'-Drh'+Ddrh')
\right.\nn
&-&\left.{\over}6r^2f'' +2Dr^2f'' +df\left((4-4d-2(1+2d-D)rh' -(d-1)r^2h'^2 -6r^2h''
+2Dr^2h''\right))\right]\nn
U&=&-{e^{-a-h}\over4(D-1)(D-2)r^2}\left[(2rf'(2+4d-2D+2drh'-Drh') -4(d^2-1+D-Dd-r^2f'')
\right.\nn
&+&f\left(4d^2-4+4D-4Dd +2(-1+2d+2d^2+D-2dD)rh'+{\over}\right.\nn
&+&\left.{\over}\left.(d-1)(1+d-D)r^2h'^2 +2r^2h'' +4dr^2h'' -2Dr^2h''\right)\right]\nn
V&=&{e^{-a-h}\over4(D-1)(D-2)r^2}\left[(D-3)rf'(4d-2rh' +2drh' +Drh') +2(-2d+2d^2 -6r^2f''
\right.\nn
&+&\left.{\over}5Dr^2f'' -D^2r^2f'')+df(4-4d-2(1+2d-D)rh' -(d-1)r^2h'^2 -6r^2h''
+2Dr^2h'')\right]\nn
X&=&-{e^{-a-h}\over4(D-1)(D-2)r^2}\left[4-4d^2-4D +4Dd+rf'(8+10d-10D-2Dd+2D^2+rh'\right.\nn
&+&5drh' -4Drh' -dDrh'+D^2rh') +6r^2f''-2Dr^2f''\nn
&+&\left.{\over}(1+d-D)f(-4+4d+2(1+2d-D)rh' +(d-1)r^2h'^2 +6r^2h'' -2Dr^2h'')\right]\nn
Y&=&{e^{-a-h}\over4(D-1)(D-2)r^2}\left[-4(d-1)(1+d-D)(-1+f) -2(4+5d+D(-5-d+D))rf'\right.\nn
&+&2(1-2d(d+1) +(2d-1)D) rfh' -(1+5d+D(-4-d+D))r^2f'h'+\nn
&+&\left.{\over}(3-(d-4)d+(d-3)D)r^2fh'^2 +2(D-3)r^2f'' -2(1+2d-D)r^2f(h'^2+h'')\right]\nn
Z&=&-{e^{-a-h}\over4(D-1)(D-2)r^2}\left[2\left(rf'(4+4d-4D+rh'+2drh'-2Drh')+\right.\right.\nn
&-&\left.{\over}2(d+d^2-D-2dD+D^2-r^2f'')\right) +(1+d-D)f\left(4(1+d-D)rh'{\over}\right.\nn
&+&\left.\left.{\over} (d-D)r^2h'^2+4(d-D+r^2h'')\right)\right]
\eea

The gravitational $\a'^3$ correction $W$ in terms
of these variables can be expressed as:
\bea
W &=& p(p-1) [ T^4 + S^4 + 4 (S^3 + T^3) Q + 3 (p-2) Q^4 + 4 Q^2 (T^2 +
S^2 + d U^2 ) + d U^4  + 4 d U^3 Q] \nn
&+& d(d-1) [X^4 + Y^4 + 4 (X^3 + Y^3 ) Z + 3 (d-2) Z^4 + 4 Z^2 (X^2 +
Y^2 + p U^2) + p U^4 + 4 p U^3 Z] \nn
&+& 2 p [T^2 S^2 + V^2 T^2 + V^2 S^2 +2 (V^2 T S + S^2 V T + T^2 V S)] \nn
&+& 2 d [X^2 Y^2 + V^2 X^2 + V^2 Y^2 +2 (V^2 X Y + Y^2 V X + X^2 V Y)] \nn
&+& 2dp [X^2 T^2 + S^2 Y^2 + U^2 (X^2 + Y^2 + T^2 + S^2)+ \nn
&+& 2 (S^2 U Y + Y^2 S U + T^2 U X + X^2 T U + U^2 T X + U^2 S Y) ]~.
\le{W}
We stress that these formulas are valid for arbitrary $p$ and $D$. In appendix \ref{explicit}
we
will give the results for specific extremal and non-extremal D3-brane solutions.

Next we compute the variation of $W$. Let us define
\be
\int\dd^Dx\sqrt{g}\,e^{-{12\over D-2}\f}\,\d W=
\int\dd^Dx\sqrt{g}\,\d g^{ij}w_{ij}
=\int\dd^Dx\sqrt{g}\,\d g_{ij}w^{ij}.
\ee
Notice that $w^{ij}=-g^{ik} g^{jl} w_{kl}$.
Explicit computation of $w_{ij}$ gives:
\be
w^{ij} = w_1^{ij} + \omega^{ij}
\ee
where
\be
w_1^{ij} =\frac{1}{2}\,e^{-{12\over D-2}\f} \left[ \left( -4
B_{mk}{}^{ln} + 4 B_{mk}^{ln}
- 3 B_m{}^{ln}{}_{k} \right) C^{mip}{}_n C^{kj}{}_{pl} -
B_{mk}{}^{li} B^{mjk}{}_l + (i \leftrightarrow j) \right]
\ee
\bea
\o^{ij} &=&\frac{1}{4} \left[ \na_n \na_m \left(D^{inmj} + D^{ijmn} -
D^{nimj} \right) \right. \nonu
&-& \frac{1}{(D-2)} \left( \na_l \na^i (d^{jl} + d^{lj}) - g^{ij}
\na_m \na_n d^{nm}  - \na^2  d^{ij}  + 2 ( R^i_m D^j{}_l{}^{lm} -
R^m_n D_m{}^{ijn}) \right. \nonu
&+& \left. \left. \frac{2}{(D-1)} \left((R^{ij} - \na^i \na^j + \na^2 g^{ij})
D_{mn}{}^{mn}  - R D_l{}^{ilj} \right) \right) + (i \leftrightarrow j)
\right]
\le{doubleu}
and we have defined
\bea
B_{ijkl} &=& C^m{}_{ijn} C^n{}_{lkm} \nonu
D_j{}^{lki} &=& e^{-{12\over D-2}\f}[\left(2 B_{jm}{}^{ni} + 3
B_j{}^{ni}{}_m -2 B_{jm}^{ni}
\right) C^{mlk}{}_n - B^{lni}{}_m C^m{}_j{}^k{}_n - B^{ln}_{mj}
C^m{}_n{}^{ki}]
-(k \leftrightarrow i) \nonu
d^{ij} &=& D_l{}^{ilj} - D^i{}_l{}^{lj}~.
\eea

\section{Explicit form of the corrections for extremal and non-extremal
D3-branes\label{explicit}}
\setcounter{equation}{0}

In this appendix we give explicit formulas for $W$ and its variation
for thermal AdS, and for the extremal and non-extremal D3-brane.
These solutions satisfy the constraint $2a_0'+h_0'=0$
(which is a necessary but not sufficient condition for supersymmetry), and we have:
\bea
e^{h_0}&=&\k_0+\left(\ell\over r\right)^{d-1}\nn
f&=&1-\left(r_0\over r\right)^{d-1}\nn
e^{\f_0}&=&g_s~.
\eea
\\
{\bf Thermal AdS$_5\times S^5$}\\
\\
The AdS limit can simply be taken by setting $\k_0=0$. The scalars $Q,\ldots,Z$ are all
given in terms of a single function $\Psi$. We get:
\bea
S &=& T = -Q = - \Psi \nn
V &=& 3 \Psi \nn
\Psi &=& \frac{r_0^4}{r^4 \ell^2}.
\eea
and
\be
W = 180 \Psi^4 ,
\ee
which agrees with the expression in \cite{GKT}.\\
\\
{\bf Extremal and non-extremal D3-branes}\\
\\
For the extremal D3-brane $r_0=0$, the result is:
\bea
Q=S&=&0\nn
T = V &=& 5 \chi \nn
U = Y &=& - \chi \nn
X &=& -4 \chi \nn
Z &=& 2 \chi \nn
\chi &=& \frac{\ell^4 \kappa_0}{r^6 }\,e^{-{5\over2}h_0} \label{wweyl}  ~.
\eea
and
\bea
W &=& 28 800 \chi^4 \label{extrW} ~.
\eea

In the non-extremal case, $W$ has a more complicated form:
\bea
W&=&{60\over r^{16}(\ell^4+\k_0r^4)^{10}}\left[3\ell^{32}r_0^{16}
+32\k_0\ell^{28}r^4r_0^{16} +219\k_0^8r^{32}r_0^{16}+\right.\nn
&+&12\k_0^7\ell^4r^{28}r_0^{12}(62r^4+23r_0^4) +2\k_0^2\ell^{24}r^8r_0^8(12r^8
-12r^4r_0^4+83r_0^8)+\nn
&+&4k_0^3\ell^{20}r^{12}r_0^8(64r^8-58r^4r_0^4 +131r_0^8) +2\k_0^6\ell^8r^{24}r_0^8 (612r^8
-4r^4r_0^4 +371r_0^8)+\nn
&+&16\k_0^5 \ell^{12}r^{20}r_0^4 (60r^{12} -14r^8r_0^4 +28r^4r_0^8 +53r_0^{12})+\nn
&+&\left. 2\k_0^4\ell^{16}r^{16} (240r^{16} -480r^{12}r_0^4 +832r^8r_0^8 -464r^4 r_0^{12}
+515r_0^{16})\right]
\eea

We also give here the variation of $W$ obtained from \eq{doubleu} for the extremal
D3-brane. One finds that the off-diagonal components are zero and the diagonal ones
are equal to
\bea
w^{tt} &=&-\frac{4800 \kappa_0^3 \ell^{12}}{r^{28} e^{10 h_0}}
\left( 56 \ell^8 - 123 \kappa_0 \ell^4 r^4 + 42 \kappa_0^2 r^8 \right)
g^{tt} \nonu
w^{aa} &=&-\frac{4800 \kappa_0^3 \ell^{12}}{r^{28} e^{10 h_0}}
\left( 56 \ell^8 - 123 \kappa_0 \ell^4 r^4 + 42 \kappa_0^2 r^8 \right)
g^{aa} \nonu
w^{rr} &=&-\frac{9600 \kappa_0^3 \ell^{12}}{r^{28} e^{10 h_0}}
\left( 7 \ell^8 - 9 \kappa_0 \ell^4 r^4 + 6 \kappa_0^2 r^8 \right)
g^{rr} \nonu
w^{mm} &=&\frac{1920 \kappa_0^3 \ell^{12}}{r^{28} e^{10 h_0}}
\left( 119 \ell^8 - 267 \kappa_0 \ell^4 r^4 + 90 \kappa_0^2 r^8 \right)
g^{mm}
\eea
where $a$ runs over the spatial worldvolume coordinates and $m$ over
the coordinates of the sphere.

\section{The most general solution of the lowest order equations \label{gene}}
\setcounter{equation}{0}

In this appendix we present the most general solution of the lowest
order equations (\ref{eaex})-(\ref{ephiex}). An analysis of this
system has also been presented in \cite{Zhou:1999nm}. Let us define
\be
\a=2 a'+ h', \qquad \b = h', \qquad \g = \f'
\ee
Consider the
following linear combinations of the equations
\bea
{1 \over 5} [(\ref{eaex})- (\ref{efex})]: &&
\a' + {9 \over r} \a + 2 \a^2 =0 \label{alpha} \\
(\ref{ephiex}): && \g' + \g (2 \a + {5 \over r}) =0 \label{gamma} \\
-{4 \over 5} [(\ref{eaex})- {9 \over 4} (\ref{efex})]: &&
\b' + {5 \over r} \b + \b^2 - {36 \over r} \a - 9 \a^2 + \g^2 =0
\label{beta}
\eea
The solution of these equations should still satisfy (\ref{ehex}).

Equations (\ref{alpha})-(\ref{gamma})-(\ref{beta}) can be integrated
by elementary means. Let us first consider the special case
\be \label{sp}
\a = 0 \quad \Rightarrow \quad h+2 a = c_1
\ee
Then we get
\bea
\g =
{4 c_2 \over r^5} \quad & \Rightarrow & \quad \f = c_3 - {c_2 \over r^4} \\
\b = -{4 c_2 \over r^5} \tan \left( c_4 - {c_2 \over r^4} \right)
\quad & \Rightarrow & h = c_5
+ \log \cos \left(c_4 - {c_2 \over r^4}  \right)
\eea
Inserting in (\ref{ehex}) we get
\be \label{inte}
c_2 e^{c_1 + c_5} = \pm l^4
\ee
Requiring that the solution is asymptotically flat
and the dilaton approaches 1 asymptotically fixes
\be
c_1=0, \qquad e^{-c_5} = \cos c_4, \qquad c_3=1.
\ee
We thus finally get the solution
\bea \label{d3ext}
ds^2 &=&
\left({\cos (c_4 - {l^4 \cos c_4 \over r^4}) \over \cos c_4}\right)^{-1/2}
(-dt^2 + d \vec{x}^2)
+ \left({\cos (c_4 - {l^4 \cos c_4 \over r^4}) \over \cos c_4}\right)^{1/2}
(dr^2 + r^2 d\O_5^2)   \nonu
\phi &=& 1 \mp {l^4 \cos c_4  \over r^4}
\eea
and  the self-dual $F_5$ is given in (\ref{sdsol}).
The $\mp$ sign in the dilaton is related to the two sign choices
in (\ref{inte}). Neither the metric nor the five form
depend on these signs. The reason is that one can change the relative
sign in $c_4 - l^4 \cos c_4/r^4$ by taking $c_4 \to - c_4$.
This does not affect the metric and the five-form because this
conbination appears inside the cosine.
The standard supersymmetric D3 brane solution is obtained by
the limit $\cos c_4 \to 0$.

Equation (\ref{alpha}) admits more general solutions than (\ref{sp}).
The most general solution of (\ref{alpha}) is
\be
\a = {4 \over (r^8 d_0 - 1) r} \quad \Rightarrow \quad
h+2 a = d_1 + \half \log \left( d_0 -{1 \over r^8} \right)
\ee
Here and in the following we assume $d_0>0$, but a similar analysis
can be done for $d_0<0$. 
Inserting the solution of $\a$ in (\ref{gamma}) and integrating we obtain
\be
\g = {d_2 r^3 \over d_0 r^8 -1} \quad \Rightarrow \quad
\f = d_3 +
{d_2 \over 8 \sqrt{d_0}} \log {\sqrt{d_0} r^4 -1 \over \sqrt{d_0} r^4 + 1}~.
\ee

Equation (\ref{beta}) becomes
\be \label{beta1}
\b' + {5 \over r} \b + \b^2 + {64 d_0 D r^6 \over (d_0 r^8 -1)^2} =0
\ee
where $D=(d_2^2 - 144 d_0)/64 d_0$. This can be solved as follows. Let us
define
\be
H=e^h, \qquad \rho = \half ({1 \over \sqrt{d_0} r^4} +1)
\ee
In terms of these variables (\ref{beta1}) becomes
\be
\pa_\rho^2 H + {D \over \r^2 (1 -\r)^2} H =0.
\ee
The most general solution of this equation is
\be
H=d_4 \r^{\a_+} (1-\r)^{\a_-}
+ d_5 \r^{\a_-} (1-\r)^{\a_+}
\ee
where $\a_{\pm} = \half (1 \pm \sqrt{1 - 4 D})$. The exponents are
real for $D \leq 1/4$. 
The case
$D=1/4$ is a special case since in this case $\a_+=\a_-$.
In this case the second independent solution involves a logarithm.
One should still impose (\ref{ehex}) which should relate
the integration constants to the scale $l$.
Asymptotic flatness fixes
$d_4 + d_5 = 2, d_1 = - \half \log d_0, d_3=1$.

\end{document}